\documentclass{aa}  
\makeatletter
\renewcommand*\aa@pageof{, page \thepage{} of \pageref*{LastPage}}
\makeatother
\usepackage{graphicx}
\usepackage{txfonts}
\usepackage{multirow}
\usepackage{subcaption}
\usepackage[switch]{lineno}
\usepackage{hyperref}
\usepackage{cleveref}
\hypersetup{
    colorlinks=true,
    linkcolor=blue,
    filecolor=magenta,      
    urlcolor=cyan,
    citecolor=blue,
}

\usepackage{graphicx,psfig,epsf}  
\usepackage{times}
\usepackage{textcomp}
\usepackage{multirow, longtable}
\usepackage{amsmath}	% Advanced maths commands
\usepackage{amssymb}	% Extra maths symbols
\usepackage{gensymb}
\usepackage{colortbl}
\usepackage{comment}

\usepackage{array,etoolbox}
\preto\tabular{\setcounter{magicrownumbers}{0}}
\newcounter{magicrownumbers}

\def\apjl{ApJ}               
\def\apjs{ApJS}               
\def\ao{Appl.~Opt.}         
           
\def\aap{A\&A}

\def\swift{{\it Swift/XRT}}
\def\nus{{\it NuSTAR}}

\def\xmm{{\it XMM-Newton}}

\def\flux{erg s$^{-1}$ cm$^{-2}$}
\def\lum{erg s$^{-1}$}

\usepackage[normalem]{ulem}

\begin{document}

\title{Detection of kilosecond hard lags in the new pulsating ULX candidate NGC 7456 ULX-1}
\titlerunning{Hard lags in NGC 7456 ULX-1}

\author{
W. Leone\inst{1}\thanks{E-mail: wladimiro.leone@inaf.it}
\and F. Pintore\inst{2}
\and C. Pinto\inst{2}
\and N. O. Pinciroli Vago\inst{3,4,10}
\and A. Sanna\inst{1}
\and A. Wolter\inst{5}
\and T. Di Salvo\inst{6}
\and R. Iaria\inst{6}
\and A. D'A\`i\inst{2}
\and A. Anitra\inst{6}
\and R. Soria \inst{7}
\and P. Esposito\inst{8}
\and F. Barra\inst{2,6}
\and E. Ambrosi\inst{2}
\and S. Caserta\inst{2,6}
\and C. Salvaggio\inst{5}
\and R. Salvaterra\inst{9}
\and G. L. Israel\inst{4}
\and S. Banerjee\inst{2}
\and M. Marelli\inst{9}
\and G. Rodriguez-Castillo\inst{2}
\and L. Burderi\inst{1}
}

\authorrunning{Leone et al.}

\institute{
Dipartimento di Fisica, Universit\`a degli Studi di Cagliari, SP Monserrato-Sestu, km 0.7, 09042 Monserrato, Italy
\and
INAF -- IASF Palermo, Via U. La Malfa 153, 90146 Palermo, Italy
\and
Department of Electronics, Information and Bioengineering, Politecnico di Milano, Via G. Ponzio 34, 20133 Milano, Italy
\and
INAF -- Osservatorio Astronomico di Roma, Via Frascati 33, 00078 Monte Porzio Catone (RM), Italy
\and
INAF -- Osservatorio Astronomico di Brera, Via Brera 28, 20121 Milano, Italy
\and
Dipartimento di Fisica e Chimica, Universit\`a degli Studi di Palermo, Via Archirafi 36, 90123 Palermo, Italy
\and
INAF-Osservatorio Astrofisico di Torino, Strada Osservatorio 20, I-10025 Pino Torinese, Italy
\and
Scuola Universitaria Superiore IUSS Pavia, Palazzo del Broletto, Piazza della Vittoria 15, 27100 Pavia, Italy
\and
INAF -- IASF Milano, Via Alfonso Corti 12, 20133 Milano, Italy
\and
AstroAI, Center for Astrophysics $|$ Harvard \& Smithsonian, 60 Garden Street, Cambridge, MA 02138, USA
}

\date{Received 21/06/2026 ; accepted 29/09/2026}

\abstract{
\textbf{Context.}
Ultraluminous X-ray sources (ULXs) are thought to be powered, in many cases, by super-Eddington accretion onto compact objects. While soft X-ray lags, commonly interpreted as reverberation in optically thick outflows, have been detected in several ULXs, hard lags remain rare and poorly understood.

\textbf{Aims.}
We investigate the temporal and energy dependence of X-ray lags in NGC~7456~ULX-1 to constrain their physical origin and probe the structure of the super-Eddington accretion flow.

\textbf{Methods.}
We analyzed the two deepest \xmm\ observations of NGC~7456~ULX-1, taken in 2018 and 2023. Hard (1--10 keV) light curves were cross-correlated with soft (0.3--1 keV) light curves using adaptive-binning techniques optimized for Poissonian low-count data. Time-resolved lag measurements were performed over consecutive 10 ks intervals, and the energy dependence of the delays was investigated. Broadband spectra were modeled using multi-component thermal and Comptonization models.

\textbf{Results.}
We detect significant hard X-ray lags in both observations, with the 1--10 keV emission delayed by $\sim10^3$ s relative to the soft band during phases of rapid flux variability. The energy-resolved analysis shows that the delays are primarily driven by the lowest-energy photons. Spectral modeling indicates a Comptonization-dominated super-Eddington flow comprising a cooler, extended outer region and a hotter, compact inner flow embedded in an optically thick wind. We interpret the observed delays as the combined effect of inward propagation of accretion-rate fluctuations and photon diffusion within the dense outflow. In this scenario, fluctuations first enhance the soft-emitting outer regions and subsequently propagate toward the hotter inner flow, where photons undergo stronger Comptonization before escaping with a kilosecond delay. The relatively small inferred inner emitting radius disfavors an intermediate-mass black hole accretor, while the observed delays imply diffusion through an extended optically thick wind.

\textbf{Conclusions.}
The sign, amplitude, and energy dependence of the observed delays disfavor a standard reverberation scenario. Instead, propagation-driven variability coupled with radiative transfer in optically thick winds appears to play a major role in shaping the timing properties of super-Eddington accretion flows.
}

\maketitle
\nolinenumbers
 
\section{Introduction}
\begin{figure*}[t!]
\center
\includegraphics[width=18.5cm]{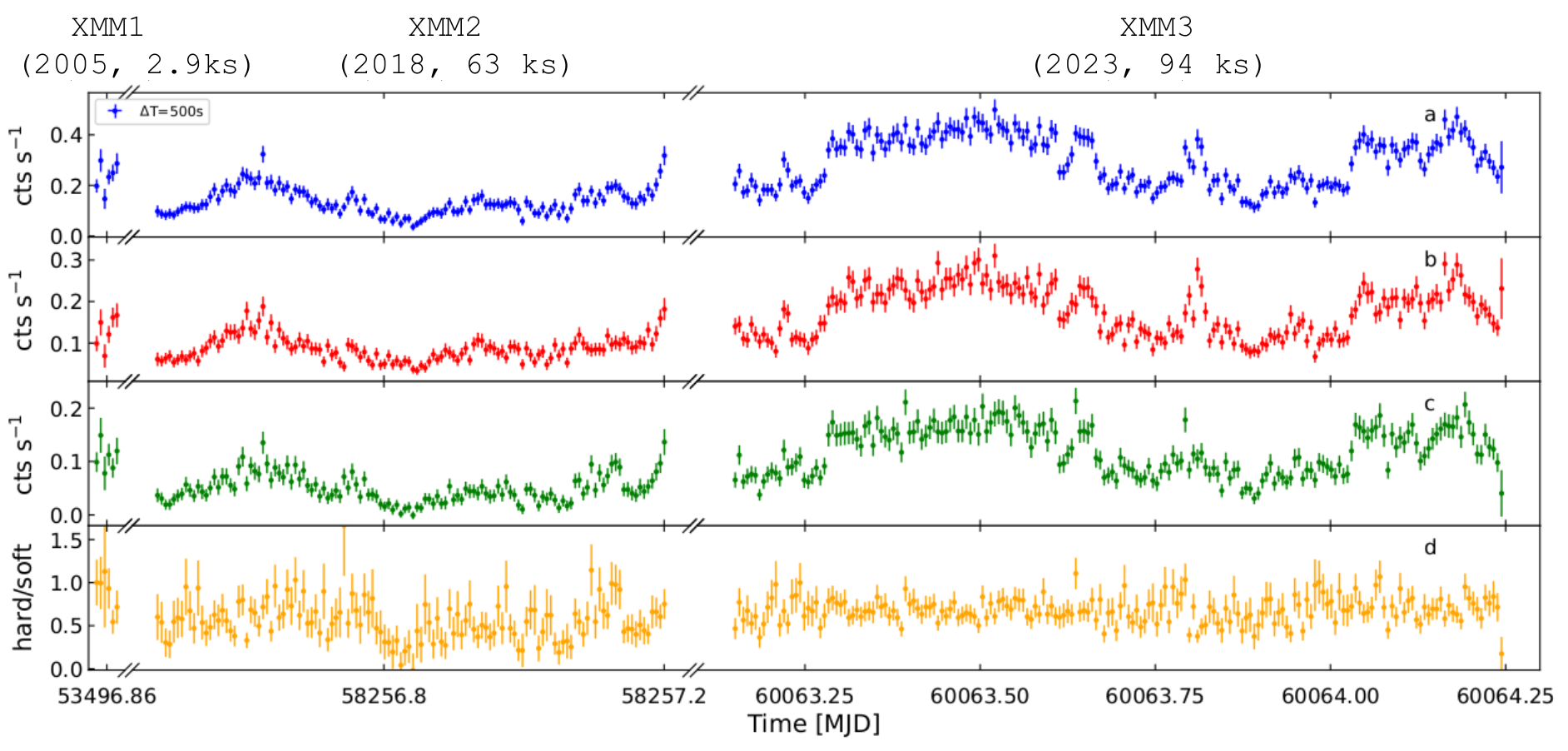}
   \caption{\xmm \, light curves (EPIC-pn+MOS) of NGC 7456 ULX-1. From top to bottom, the panels show (a) the total 0.3--10 keV band, (b) the soft 0.3--1 keV band, (c) the hard 1--10 keV band, and (d) the corresponding hardness ratio. The three segments correspond to the 2005, 2018, and 2023 \textit{XMM-Newton} observations, with net exposures of 2.9, 63, and 94 ks, respectively. The temporal gaps between the observations, which span several years, are indicated by broken time-axis. The 2005 observation is shown for completeness but it is not used in the lag analysis because of its short exposure and high background level.}
   \label{lcall}
\end{figure*}
\label{introduction}
Ultra-luminous X-ray sources (ULXs) are accreting compact objects whose extreme luminosities in the X-ray band ($L_{x} > 10^{39} \mathrm{erg/s}$) can be explained by two main scenarios: super-Eddington accretion onto stellar-mass black holes (BHs) or neutron stars (NSs), or standard sub-Eddington accretion onto intermediate-mass black holes (IMBHs).
However, the results obtained in recent decades favor the stellar-mass compact-object scenario with super-Eddington accretion rates \\ \\ \citep[e.g.][for recent reviews]{kaaret17, king23, pinto23}. A result was the discovery of seven pulsating ULXs (PULXs; \citealt{bachetti13, israel16a, israel16b, fuerst16a, carpano18, Sathyaprakash_2019, rodriguez19, ducci25}) with pulsation periods in the range $\sim$0.5-20\, s, and a strong PULX candidate in NGC 4559 X7 \citep{pintore21, pintore25}, which clearly indicates the presence of NSs in ULXs. Pulsations are detected in PULXs only in high-quality \xmm\ or \nus\ observations, because the pulsed fraction at soft X-ray (0.3-1 keV) energies is generally low, typically $\lesssim10\%$. Although the known ULXs are very numerous ($\sim 2000$, \citealt{walton22}), the actual number of ULXs with sufficient counts to enable pulse detection is significantly smaller ($<40$). This implies that the fraction of NSs in ULXs is at least $\sim20-30\%$, which is a non-negligible fraction. It is now a matter of debate whether the ULX population is mainly composed of NSs \citep[e.g.][]{pintore17, koliopanos17, walton18} or of a mixed bag of NSs and stellar mass black holes (BHs). It is therefore crucial to extend population studies by focusing efforts on constraining the nature of the compact object in as many ULX systems as possible. NSs can be uniquely identified through the detection of coherent pulsations. However, the absence of detected pulsations does not necessarily rule out a neutron-star accretor. In ULXs, pulsation searches are often limited by photon statistics, and possible dilution or reprocessing of the pulsed signal in the accretion flow and outflow. Moreover, pulsations in confirmed PULXs are known to be transient, so that observations of the same target with comparable counting statistics or spectral/luminosity properties may show pulsations in some epochs but not in others \citep[e.g.][]{belfiore24,imbrogno24}. It is therefore difficult to identify a NS in a ULX, especially when the number of high-quality observations is limited.

A way to increase the sample of candidate NSs in ULXs could be to identify properties more commonly found in PULXs \citep{pintore17, pinciroli2025hunt}.  In detail, most PULXs show large flux variations (as high as several orders of magnitude) on weekly-to-monthly timescales, whose nature has often been associated with the onset of a propeller regime \citep[e.g.][]{fuerst21, fuerst23, belfiore24}. Furthermore, PULXs have been observed during flare-like activity on timescales of minutes to hours, the origin of which remains poorly understood. It has been found that the presence of quasi-periodic oscillations in the power spectral density is often coupled to pulsation switch-offs \citep[e.g.][]{imbrogno24}. Another common characteristic of the PULXs is their super-orbital periodicity, generally on timescales of tens to hundreds of days \citep[e.g.][]{fuerst18, imbrogno24}. Finally, PULX spectra are generally harder than those of the overall ULX population, often matching the hard-ultraluminous regime \citep[e.g.][]{sutton13,pintore17}. This suggests that the emission from the hotter inner regions of the accretion flow may be more directly visible in these systems.
    
ULX spectra, and in particular those of the PULXs, are usually represented in terms of three spectral components: two have a thermal nature and modeled with (modified) blackbodies having temperatures of 0.1-0.5 keV and $\sim1$ keV, and one which is a hard, non-thermal component described by a cut-off power law\\ \citep{walton18}. The two thermal components are usually associated with either a cold, outer disk and the photosphere of powerful outflows that are expected to be ejected at super-Eddington rates \citep{poutanen07}. The hard component is instead generally associated with emission from the accretion column near the magnetic poles of the NS, while the accretion curtain may contribute to the hotter of the thermal components below $\sim$10~keV \citep[][]{walton18b}.

Furthermore, some ULXs, particularly those exhibiting softer X-ray spectra (i.e. soft-ultraluminous X-ray sources, \citealt{sutton13}), present clear evidence of features in both absorption and emission, whose nature is uniquely associated with outflows ejected by the disk \citep[e.g.][]{pinto16, kosec18a, kosec21, pinto21}. 
\begin{table}
\caption{Log of the \xmm\ observations used in this work.}
\scalebox{0.92}{
\label{list_of_log}
\begin{tabular}{lccc}
\hline
 Obs.ID & Start & Stop & Exp.$^a$ \\
 & \multicolumn{2}{c}{[YYYY-MM-DD hh:mm (UTC)]} & [ks] \\
\hline
 0303560701 & 2005-05-06T20:05 & 2005-05-06T22:25 & 2.9 \\
 0824450401 & 2018-05-18T11:24 & 2018-05-19T12:10 & 63 \\
 0901330101 & 2023-04-28T23:50 & 2023-04-30T08:49 & 94 \\
\hline
\end{tabular}
}
\footnotesize{\\ $^a$ Net exposure after screening and background subtraction in EPIC-pn;}
\end{table}
Fast, optically thick outflows are likely responsible for the hundreds-of-seconds X-ray time lags observed in some bright and highly variable ULXs, where the delays are generally interpreted as signatures of reprocessing and reverberation within the wind photosphere \citep{demarco13,pinto17,kara20}. These lags are soft, i.e., the soft emission lags the hard one, and are explained as the reprocessing of hard photons emitted by the innermost regions of the accretion flow from the funnel of the outflows \citep[e.g.][]{demarco13, pinto17, kara20}. Hard lags, i.e. delays of the hard emission with respect to the soft one, are commonly observed in accreting black holes, from black-hole X-ray binaries to active galactic nuclei, and are usually interpreted in terms of inward propagation of accretion-rate fluctuations through the accretion flow \citep[][]{Kotov_2001,arevalo06,uttley14}. To date, hard lags in ULXs have been reported only in NGC~4559~X7 \citep{pintore21} and NGC~5408~X-1 \citep{hernandez-garcia15}, on timescales of hundreds of seconds. In particular, NGC~4559~X7 is a robust PULX candidate \citep{pintore25} showing marked short-term temporal and spectral variability, identified through a search for PULX-like characteristics. Motivated by these findings, we performed a targeted search for hard X-ray lags in NGC~7456~ULX-1, investigating whether similar timing signatures are also present in this source.

\section{NGC 7456 ULX-1}

NGC~7456 is a spiral galaxy at a distance of $\sim$15.7~Mpc \citep{tully16}. It hosts at least four ULXs \citep{walton11,pintore20} and an additional candidate ULX (ULX-5; \citealt{pintore20}), whose transient nature and physical origin remain uncertain. Two sources, ULX-1 and ULX-2, are the brightest persistent objects in the galaxy, with average bolometric X-ray luminosities well above $5\times10^{39}$~\lum. The other two sources, ULX-3 and ULX-4, exhibit flux variations of at least one order of magnitude and have been proposed as candidate transient ULXs.
The galaxy was monitored by \swift\ in 2020 and 2022, and observed three times 
by \xmm. In particular, during a 2018 \xmm\ observation, ULX-1 was caught during short-term episodes of variability, with the flux varying by one order of magnitude on timescales of minutes. This represents one of the largest amplitudes of short-term X-ray variability observed in a ULX, with a fractional variability exceeding 40\% on timescales of only $\sim$500 s \citep{sutton13,middleton15a,pintore20}, without any significant accompanying spectral variability. ULX-1 can be classified as a soft-ultraluminous ULX, as defined by \citep{sutton13}. Spectra can be modeled with two thermal components  \citep{pintore20} having temperatures of $\sim0.2$ keV and $\sim0.6-0.7$ keV, plus at least one absorption line at $\sim0.7$ keV (likely O VII-VIII blueshifted lines), which suggests the presence of powerful winds in the source as it is expected for super-Eddington accretion. 
Here, we extend our analysis on ULX-1 by exploiting a new \xmm\ observation taken in 2023, in which the source is found to be at a higher flux than in 2018 but with less pronounced short-term temporal variability.

\begin{figure}[t!]
\center
\includegraphics[width=9.0cm]{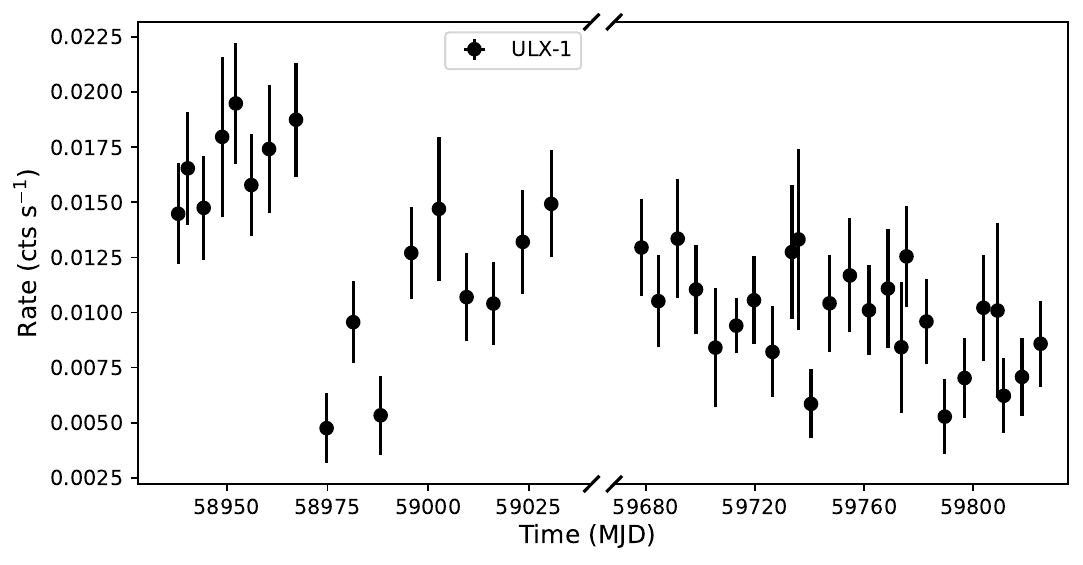}
\includegraphics[width=9.0cm]{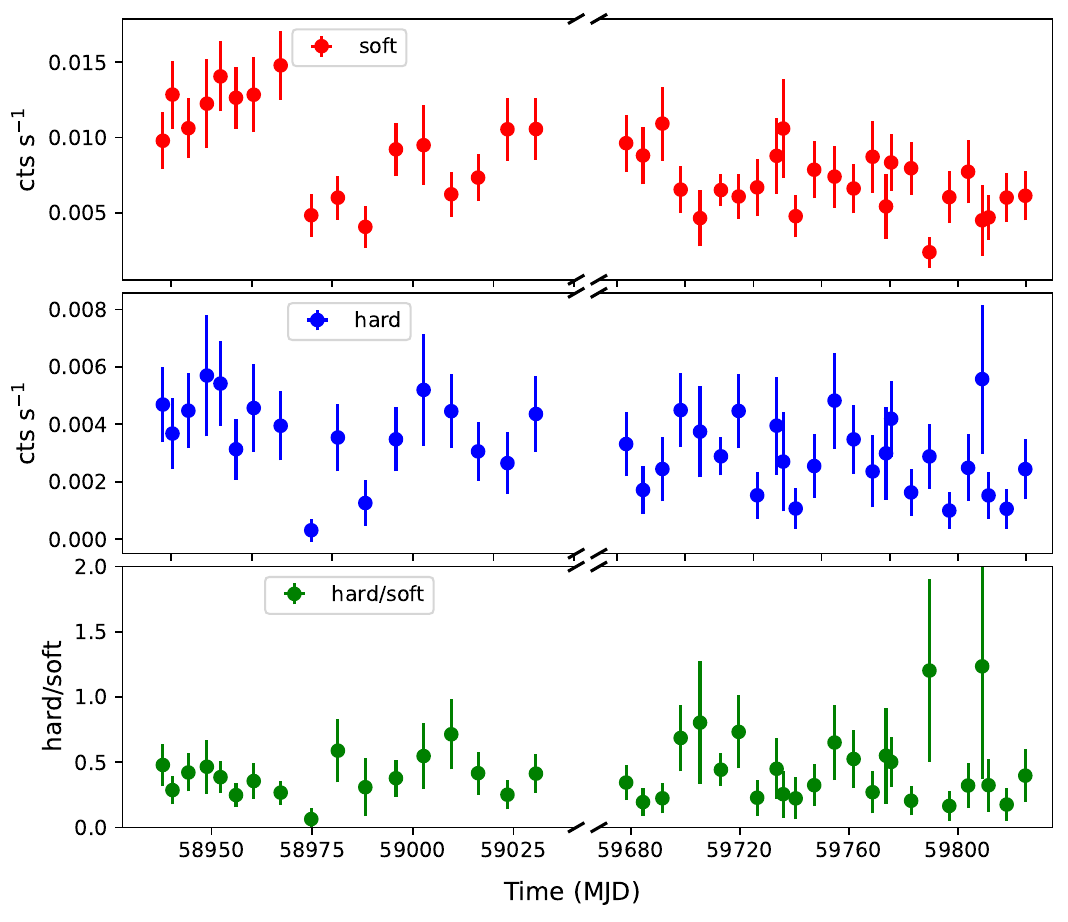}
   \caption{\textit{top}: Swift/XRT long-term light curve of ULX-1 in the total 0.3--10 keV energy band. \textit{bottom}: Soft (0.3--1.0 keV, red points) and hard (1.0--10 keV, blue points) bands and the corresponding hardness ratio (green points). Data are binned in order to have at least 15 counts per bin and a significance of at least 1.5$\sigma$. The light curve was produced using the online tool of \citet{evans09}.}
   \label{ulx_swift_hardness}
\end{figure}

\section{Data reduction}
\label{data_reduction}

{\bf XMM-Newton.} We analyzed all the available \xmm\ observations, i.e. ObsID: 0303560701 (PI: D. Rosa-Gonzalez, XMM1 hereafter), 0824450401 (PI: G. Israel, XMM2 hereafter) and 0901330101 (PI: F. Pintore, XMM3 hereafter), which were taken in 2005, 2018 and 2023, respectively (Table~\ref{list_of_log}). 

 We reduced the \xmm\ data using the SAS v21.0.0 with the most recent calibration files. We selected the data from both EPIC-pn and EPIC-MOS using the {\sc epproc} and {\sc emproc} tasks. Events were filtered with standard criteria, i.e. FLAG=0 and PATTERN$\leq$4 (12) for pn (MOS). We obtained good time intervals by removing all intervals with a flaring particle background. In particular, light curves and spectra were extracted in the following time intervals: [231797254:231800156]s, [643029762:643092614]s and [799126347:799220865]s from the \xmm\ reference time, resulting in a net exposure time of 2.9 ks, 63 ks and 94 ks for XMM1, XMM2, and XMM3, respectively. 
Source and background events were selected from circular regions of radii 30'' and 45'', with the latter taken from a source-free region near ULX-1 on the same CCD. The photon times of arrival (ToAs) were reported to the barycenter of the Solar System using the task {\sc barycen}.
X-ray spectra were grouped using the {\sc heasoft} tool {\sc grppha} to obtain at least 25 counts per energy bin. Source and background light curves were extracted with the {\sc SAS} task {\sc evselect} and then corrected with the task {\sc epiclccorr}. The final result is shown in Fig~\ref{lcall}.

{\bf Swift/XRT.} We also analyzed all the available \swift\ observations, which were taken during two monitoring campaigns with a bi-weekly cadence in 2020 and 2022. Data were reduced with the online tool \\ \url{https://www.swift.ac.uk/user_objects/} of \citet{evans09} and used to build the long-term ULX-1 light curve in the soft (0.3-1 keV), hard (1-10 keV), and in the full (0.3--10 keV) X-ray bands.

\vspace{-0.4cm}
\section{Timing analysis}

\begin{figure}[t!]
    \centering
    \includegraphics[width=\linewidth]{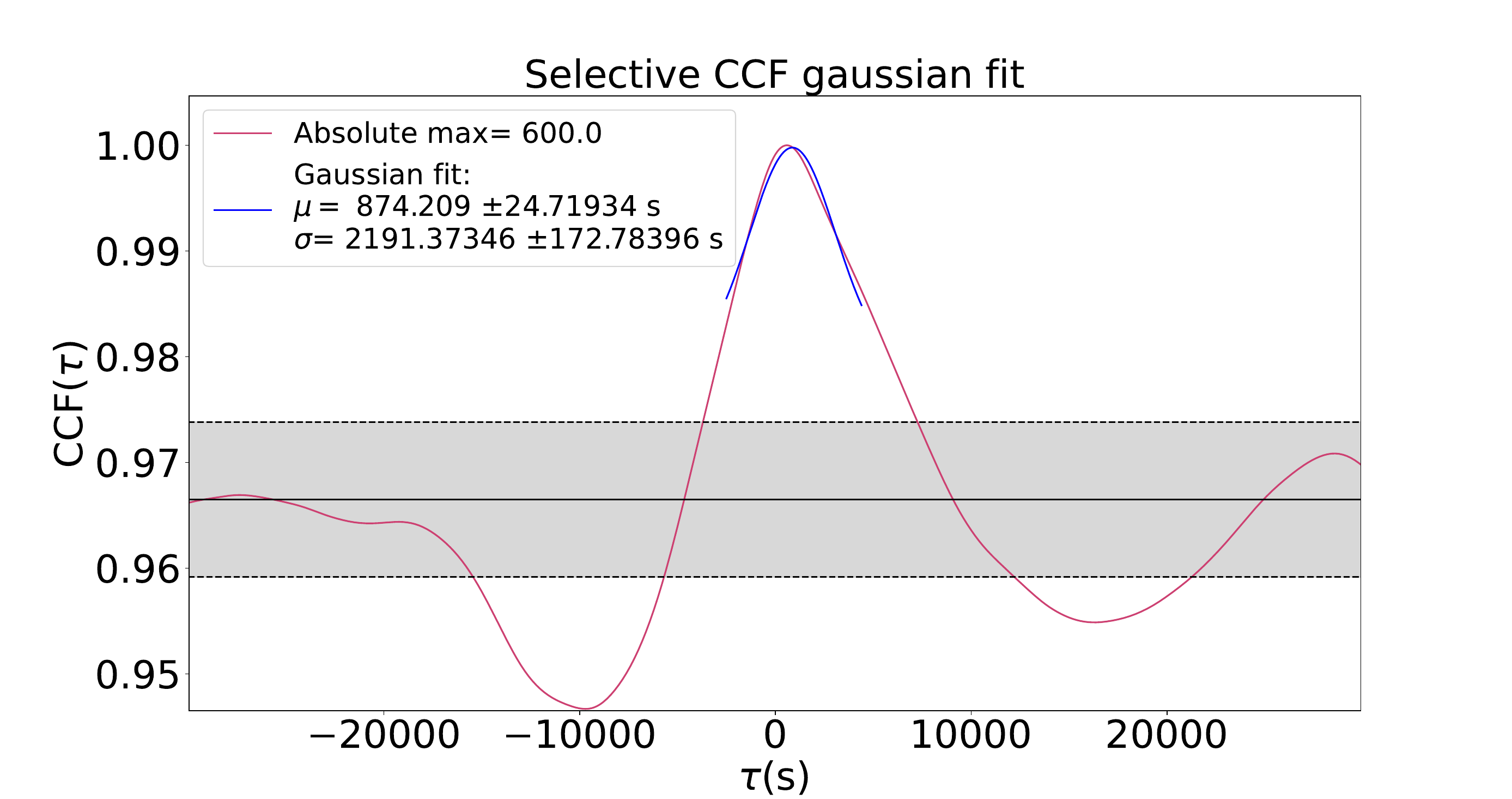}
    \caption{
Cross--correlation of the 1--10~keV (hard) band with respect to the
0.3--1~keV (soft) band computed with the MDP procedure
(adaptive bin size, $n=100$ photons per bin).
The CCF is evaluated over the full, GTI-corrected XMM2 observation;
a Gaussian fit, applied to the entire CCF profile, is overplotted,
and the corresponding best-fit parameters are quoted.
The orange area indicates the 68\% confidence region of the CCF profile.
}
    \label{fig:CCF_total_XMM2}
\end{figure}

To investigate the super-Eddington accretion scenario and the geometry of the emitting regions, we compare the timing properties of the hard (1--10 keV) and soft (0.3--1 keV) X-ray bands, following \citet{pinto23}. This choice is motivated by the typical curved spectra of ULXs, in which the transition between the cooler and hotter thermal components occurs around $\sim 1$ keV. Thus, the soft band mainly probes the cooler, more extended emission region, whereas the hard band traces the hotter inner flow and/or Comptonized radiation.
    
\subsection{Long-term analysis (days-to-months)}
\label{long_term}

The observations of ULX-1 span almost the last 20 years, although coverage is very sparse, with many gaps during this interval. In addition to the three \xmm\ observations, \swift\ monitored ULX-1 in 2020 and 2022, allowing us to build the most continuous light curve for this source (Fig.~\ref{ulx_swift_hardness}). 
\newline
\newline
The monitoring showed that ULX-1 was persistently observed in accretion, with clear evidence of flux variability in the 0.3--10 keV energy band of approximately a factor of 5. In addition, a marked flux drop was observed at around MJD 59000 (Fig. \ref{ulx_swift_hardness}-top). The evolution of the emission in the 0.3--1 keV and 1--10 keV energy bands indicates that the components responsible for the emission in these bands may vary simultaneously (Fig.~\ref{ulx_swift_hardness}-right), confirming the results already found for the \xmm\ observations in \citet{pintore20}.
The alternative possibility explored in this work is that both the soft and hard emission components vary in response to fluctuations in the accretion flow, while their intrinsic spectral evolution is largely hidden by reprocessing in an optically thick wind. Multiple scatterings redistribute photon energies and blur the temporal response of the different emitting regions, thereby damping the observable spectral variability and producing only modest changes in the hardness ratio.
 \begin{figure}[t!]
    \centering
    \includegraphics[width=\linewidth]{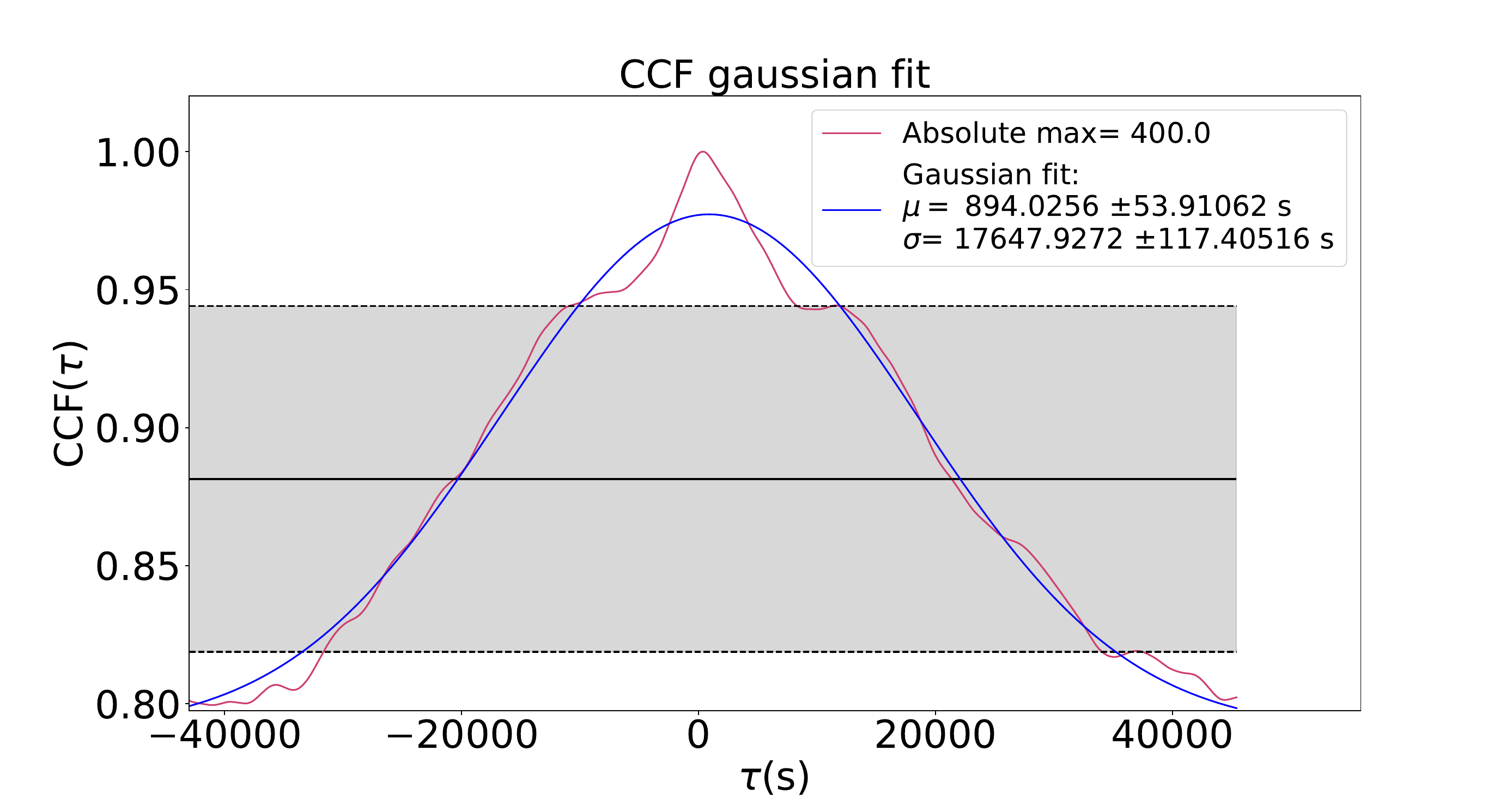}
    \caption{
Cross--correlation of the 1--10~keV (hard) band with respect to the
0.3--1~keV (soft) band for the full, GTI-corrected XMM3 observation,
computed with the same MDP setup as in Fig.~\ref{fig:CCF_total_XMM2}.
In this case, the Gaussian fit is applied to the entire CCF profile.
}
    \label{fig:CCF_total_XMM3}
\end{figure}

Hints of a spectral variation are also detected (Fig.~\ref{ulx_swift_hardness}-bottom), although the low counts of the \swift\ prevent us from meaningful spectral modeling. We note that the soft emission exhibits a maximum jump in flux of about a factor of 6 (without accounting for uncertainties), whereas the hard emission is as high as a factor of $\sim$20. It is clear, therefore, that the two bands experience significant variations, but most of the flux variability is carried by the hard one. This suggests that the spectral components responsible for the emission in the two bands may not be fully connected.
\subsection{Short-term variability (minutes-to-hours)}
\label{short-term_variability}

\begin{figure}[t!]
    \centering
    \includegraphics[width=0.9\linewidth]{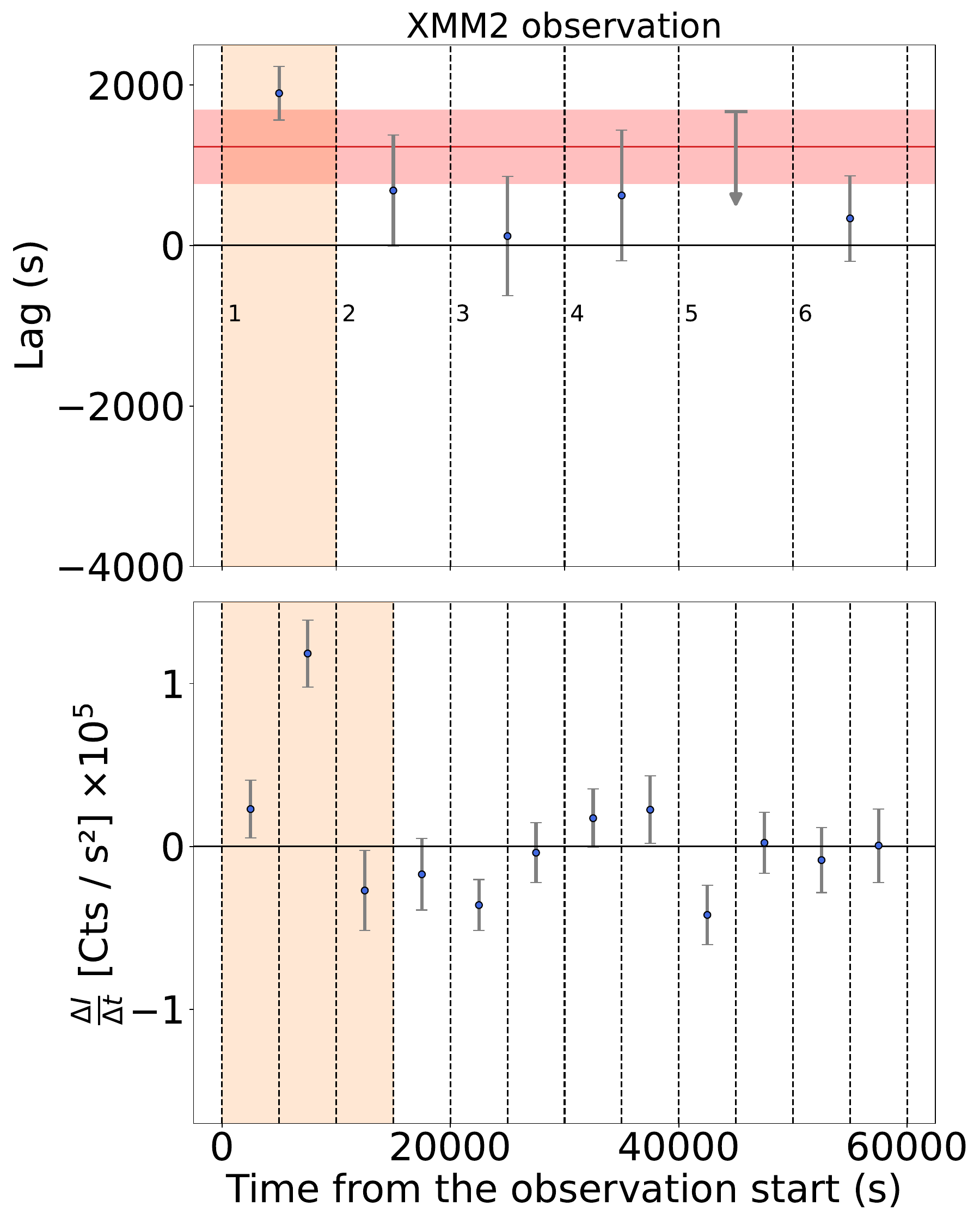}
    \caption{
    Top: Lags as a function of time, evaluated in $10\,\mathrm{ks}$
    windows for the XMM2 observation. Errors are at $1\sigma$.
    Positive lags indicate that the soft band leads the hard band
    (i.e., a hard lag). The red horizontal line indicates the lag measured
    from the full observation, while the shaded red region denotes the
    corresponding $1\sigma$ uncertainty obtained from the Monte Carlo MDP
    analysis of the entire exposure. This uncertainty is independent of the
    time-resolved lag measurements and should not be interpreted as the
    dispersion or average of the individual points.
    Bottom: Temporal evolution of the count-rate derivative
    (i.e., rate-curve slope vs.\ time) in contiguous $5\,\mathrm{ks}$ windows;
    errors correspond to the 68\% confidence intervals.
    }
    \label{fig:lc_cross_corr_xmm2}
\end{figure}

{\bf ULX-1:} As shown in Fig. \ref{lcall}, all \xmm\ observations give strong indications that ULX-1 is highly variable in flux on short-timescales (i.e., on minutes to hours). We estimated the fractional variability F$_{var}$ \citep[e.g.][]{vaughan03} in the three data sets over comparable time scales for XMM2 and XMM3 (100s-T$_{obs}$, where T$_{obs}$ is the observation exposure) and smaller for XMM1: we found that F$_{var}$ in the energy band 0.3--10 keV is $<30$\% (3$\sigma$ upper limit), $37\pm1.5$\% and $32\pm1$\% for XMM1, XMM2 and XMM3, respectively. We also estimated F$_{var}$ in the 0.3--1 keV and 1--10 keV energy bands. We obtained $<35$\% (3$\sigma$ upper limit), $32\pm2$\% and $30\pm1$\% in the soft band, and $<40$\% (3$\sigma$ upper limit), $47\pm3$\% and $33\pm1$\% in the hard band, for XMM1, XMM2 and XMM3, respectively.

\begin{figure}[t!]
    \centering
    \includegraphics[width=0.9\linewidth]{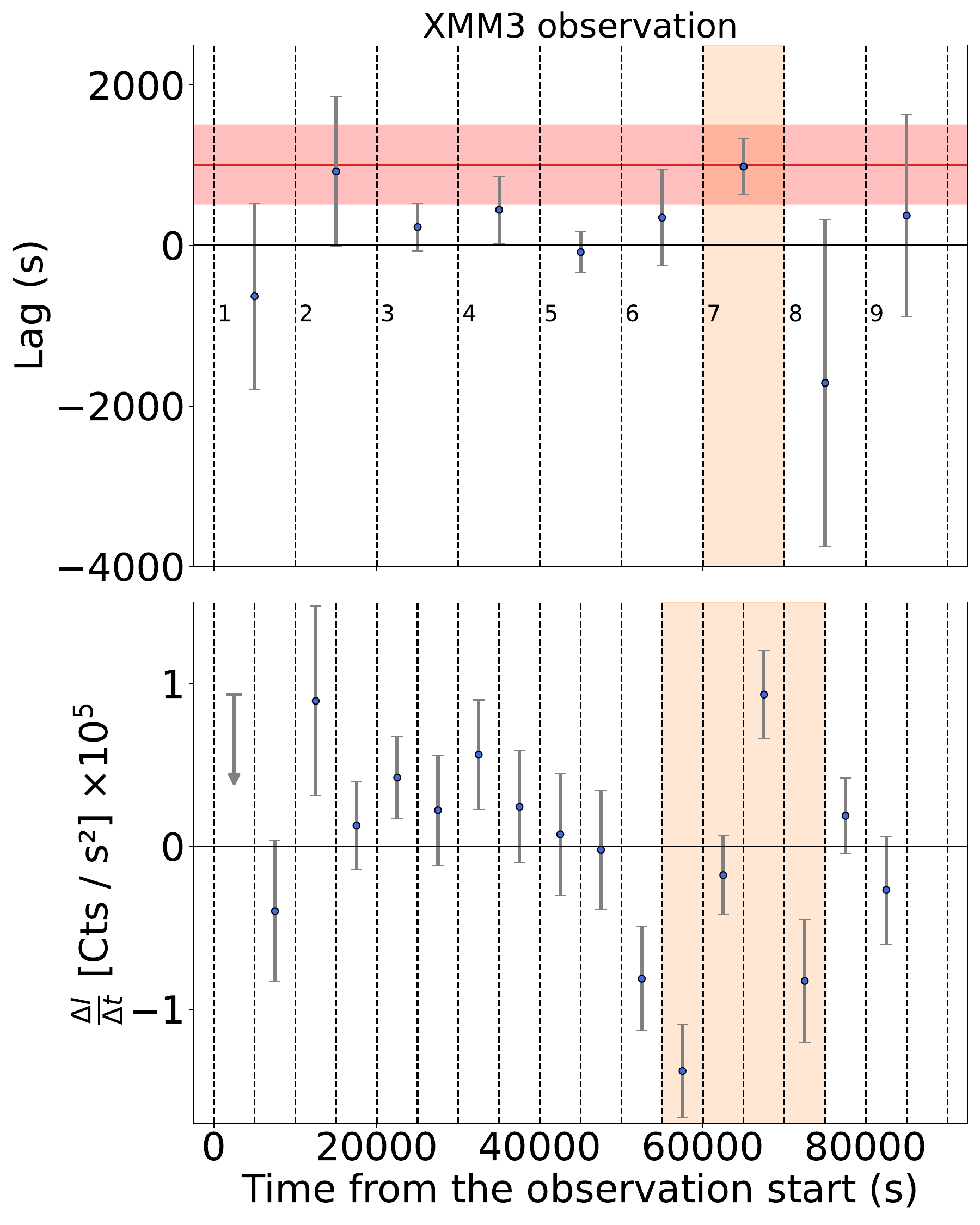}
    \caption{
    Same as Fig.~\ref{fig:lc_cross_corr_xmm2}, but for the XMM3 observation.
    }
    \label{fig:lc_cross_corr_xmm3}
\end{figure}

Furthermore, we quantified the source variability in successive, non-overlapping 5 ks intervals in the 0.3--10 keV band using the local count-rate derivative, ($\Delta I/\Delta t$), estimated from a linear fit to the count rate as a function of time within each interval. We note that the choice of a 5 ks timescale represents a compromise between temporal resolution and counting statistics. The analysis was performed separately for XMM2 and XMM3 using only GTI-filtered data.
To avoid introducing artificial variability, we considered only 5 ks intervals fully contained within individual good-time intervals, excluding segments that would otherwise cross data gaps. Within each 5 ks segment, the light curve was rebinned to a fixed time resolution of $\Delta t = 250$ s.
Within these intervals, the number of detected photons per bin is typically large enough for Gaussian statistics to provide a good approximation to the Poisson distribution. Under this condition, the per–bin rate uncertainty can be written as:\vspace{-0.2cm}
\[
\sigma_{\rm rate} \;\simeq\; \frac{\sqrt{N}}{\Delta t}
\;=\; \frac{\mathrm{rate}}{\sqrt{N}},
\]
where $N$ is the number of counts in each $\Delta t$.

\subsection{Time lag estimation}

We searched for time lags by analyzing the longest \xmm\ observations (XMM2 and XMM3), which have the highest count statistics and the highest short-term variability, and excluded XMM1 due to the high particle background throughout the full observation. Lags in ULXs are more readily identified in light curves exhibiting flares or dips, when temporal signatures of features across energy bands are more strongly correlated. 

Here, we adopted the cross-correlation function (CCF) formalism described in \cite{Leone2025}: the technique enables robust delay estimation directly from photon ToA lists by constructing continuous light curves through adaptive bin-size methods and by quantifying Poisson-induced uncertainties via Monte Carlo simulations, thus overcoming the limitations of traditional fixed binning in the low-count regime. An extensive description of the methodology is in Appendix \ref{appendix:adaptive_binning}.

In particular, we constructed the light curves using an adaptive binning based on a fixed number of counts per bin, adopting \(n=100\) photons for both XMM2 and XMM3. This choice sets the relative Poisson uncertainty of the count rates in each time bin to $1/\sqrt{n}=10\%$. 
Uncertainties on the derived lags are obtained with the Modified Double Pool (MDP) procedure described in \cite{Leone2025}.
For clarity, Figs.~\ref{fig:CCF_total_XMM2} and Fig.~\ref{fig:CCF_total_XMM3} present two representative CCF realizations from the MDP Monte Carlo procedure, together with examples of the two possible fitting windows: the significant CCF peak, when uniquely identified, or the entire CCF profile otherwise.
We note that the uncertainties quoted in the legend of Fig.~\ref{fig:CCF_total_XMM2} and Fig.~\ref{fig:CCF_total_XMM3} for the Gaussian centroids are formal fit uncertainties for the representative CCFs shown here. The reported uncertainties should not be confused with the ones on the final lag measurements, which are derived from the MDP Monte Carlo distributions, as described in Appendix \ref{appendix:adaptive_binning}.\\
By our sign convention, when cross-correlating the hard-photon light curve with the soft-photon one, a negative/positive lag means soft-photons lag/anticipate the hard-photons, i.e., soft-lag/hard-lag.
We calculated the delay between the hard (1--10 keV) and soft (0.3--1 keV) energy bands. These are the ranges over which each ULX-1 spectral component, identified in the average spectrum, dominates (see Section~\ref{section:spec_analysis}). 
The total lags between the two bands evaluated for XMM2 and XMM3 are $\tau_{tot,XMM2}=1230\pm 460$\,s and $\tau_{tot,XMM3}=1010\pm 500$\,s, both reported in Fig.~\ref{fig:lc_cross_corr_xmm2} and Fig.~\ref{fig:lc_cross_corr_xmm3} as shaded red regions. In both cases, the delay is positive, implying a
`hard' lag (i.e. hard emission lags the soft one), in contrast with the `soft' lags (i.e. soft emission lags the hard one) usually found in ULXs. 

\begin{figure}[t!]
    \centering
    \includegraphics[width=\linewidth]{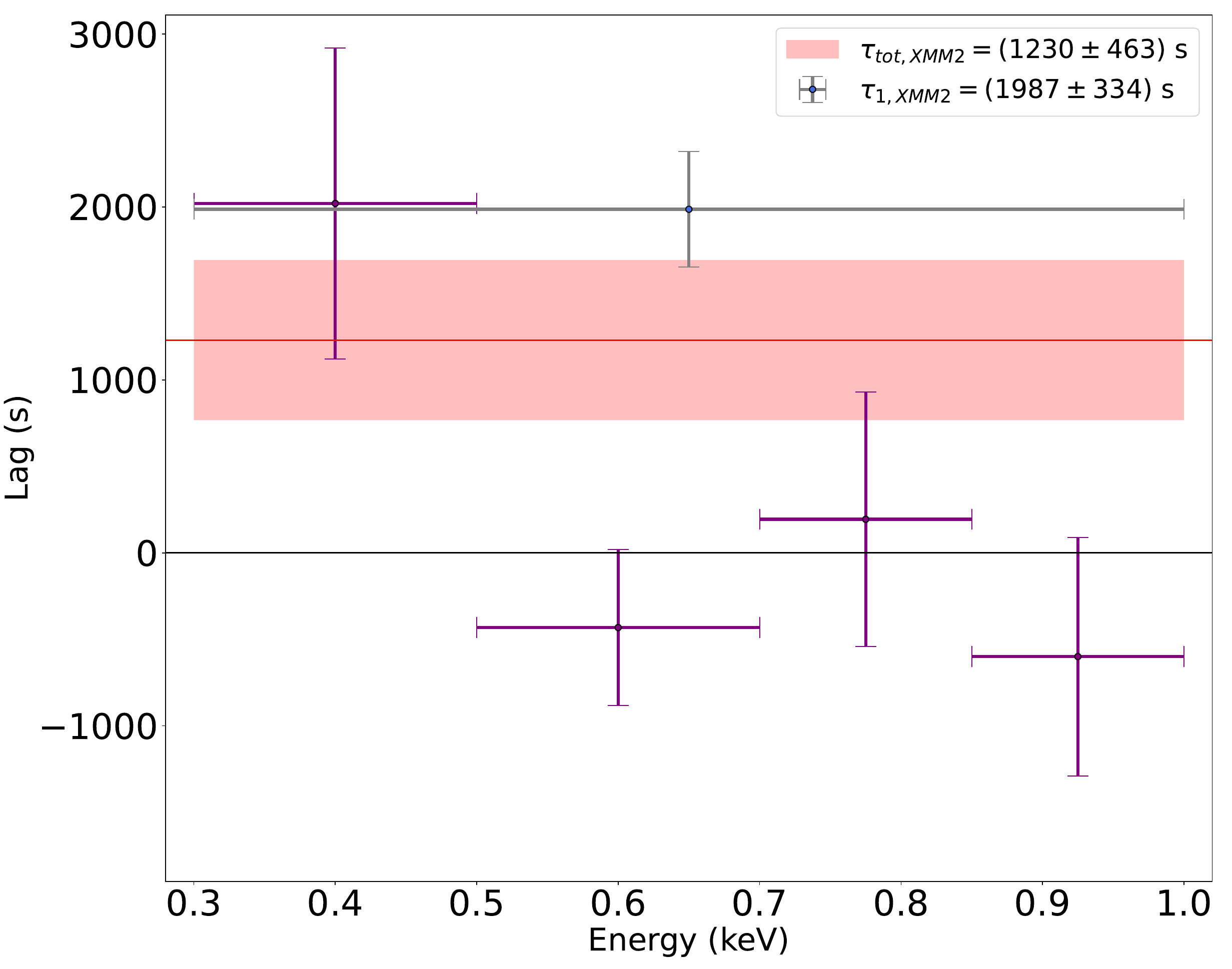}
    \caption{
    Energy-resolved time lags for XMM2-1 (0--10 ks).
    The red band represents the lag measured over the full observation interval,
    while the gray point indicates the lag computed in the time window showing
    the strongest correlation signal (see Fig.~\ref{fig:lc_cross_corr_xmm2}).
    Vertical error bars correspond to $1\sigma$ uncertainties on the lag
    measurement, and horizontal bars indicate the energy range of each soft
    band used to compute the lag of the 1--10~keV band with respect to the soft
    sub-bands reported here.
    }
    \label{fig:energy_resolved_lag_XMM2}
\end{figure}

The methodology introduced in \citet{Leone2025} is particularly effective in the low-counts regime, therefore, we could investigate the time dependence of the positive lags in ULX-1 by dividing each observation into time intervals of 10~ks. Owing to the limited photon statistics, 10 ks was adopted as the shortest timescale that ensures reliable CCF measurements while preserving adequate temporal resolution. As shown in the top panel of Fig.~\ref{fig:lc_cross_corr_xmm2} and Fig.~\ref{fig:lc_cross_corr_xmm3}, we detected two significant ($\sim 5.9\sigma$ and $\sim2.8\sigma$ respectively) hard lags in the first ([0-10] ks) time window for XMM2, with magnitude $\tau_{1,XMM2}=1987\pm 334$\,s, and in the seventh ([60-70] ks) window for XMM3, with magnitude of $\tau_{7,XMM3}=982\pm 345$\,s. The significance of the lag is evaluated in each 10 ks interval as $\tau/\sigma_{\tau}$, where $\tau$ and $\sigma_{\tau}$ (68 \% C.L.) are the mean and standard deviation of the MDP Monte Carlo lag distribution. These values represent local significances and do not include a correction for the number of 10 ks intervals examined.
Interestingly, the significant hard lags detected in the two 10 ks intervals highlighted in Fig.~\ref{fig:lc_cross_corr_xmm2} and Fig.~\ref{fig:lc_cross_corr_xmm3}-top occur simultaneously with pronounced variations in the local count-rate derivative, ($\Delta I/\Delta t$), shown in Fig.~\ref{fig:lc_cross_corr_xmm2} and Fig.~\ref{fig:lc_cross_corr_xmm3}-bottom. The physical implications of this behavior are discussed in Sect.~\ref{discussion}. We note that the lag measurements across individual intervals vary, consistent with zero in intervals without pronounced changes in $\Delta I/\Delta t$. This may indicate that statistically significant lags are more easily detected during phases of enhanced short-term variability, when the correlation between the soft and hard bands is stronger. However, these intervals also coincide with higher count-rate phases, and hence with better photon statistics. Therefore, the increased lag significance may reflect both improved counting statistics and a stronger physical coupling between the soft and hard-band variability. However, the average lags measured over the full XMM2 and XMM3 observations are remarkably similar, and the weighted mean of the lag measurements obtained from the two individual intervals is consistent with these full-observation values. This suggests that a single isolated episode should not drive the hard-lag signal, but rather represents a recurrent feature of the variability pattern observed in NGC~7456~ULX-1.

As an independent check, we reproduced the analysis with the HEASoft task 
{\sc crosscor}\footnote{See \href{https://heasarc.gsfc.nasa.gov/docs/software/lheasoft/help/crosscor.html}{HEASOFT crosscor}}, 
reaching consistent results when using the same light curves presented in this work. 
In particular, when analyzing the full observations, {\sc crosscor} recovered lag values consistent with those obtained with the MDP method, although with larger uncertainties and, consequently, lower statistical significance. 
This is expected, since the adaptive-binning approach in Appendix \ref{appendix:adaptive_binning} was specifically developed to optimize delay 
measurements in the low-count regime and to mitigate the limitations of 
fixed-bin-size cross-correlation methods, discussed in \citep{Leone2025}. The difference becomes more pronounced in the 10~ks time windows, where the reduced photon statistics 
prevent {\sc crosscor} from significantly constraining the lag, yielding 
only upper limits consistent with the MDP measurements, whereas the MDP 
method still provides statistically meaningful lag estimates. We also measured the magnitude-squared coherence between the hard and soft X-ray light curves using the Welch-averaged periodogram \citep{Welch}. This provides an independent diagnostic of the reliability of the lag measurement, since a meaningful time delay is expected only when the two bands show correlated variability. The coherence reaches a maximum value of $\sim$30\% at $f \approx 4 \times 10^{-4}$ Hz and is highest at low frequencies ($\lesssim$ mHz), corresponding to the timescales associated with the measured lags.

Taken together, the independent replication of the signal and the enhanced low-frequency coherence support the robustness of the measured delay, indicating that it originates from a genuine physical process rather than from statistical fluctuations.

\begin{figure}[t!]
    \centering
    \includegraphics[width=\linewidth]{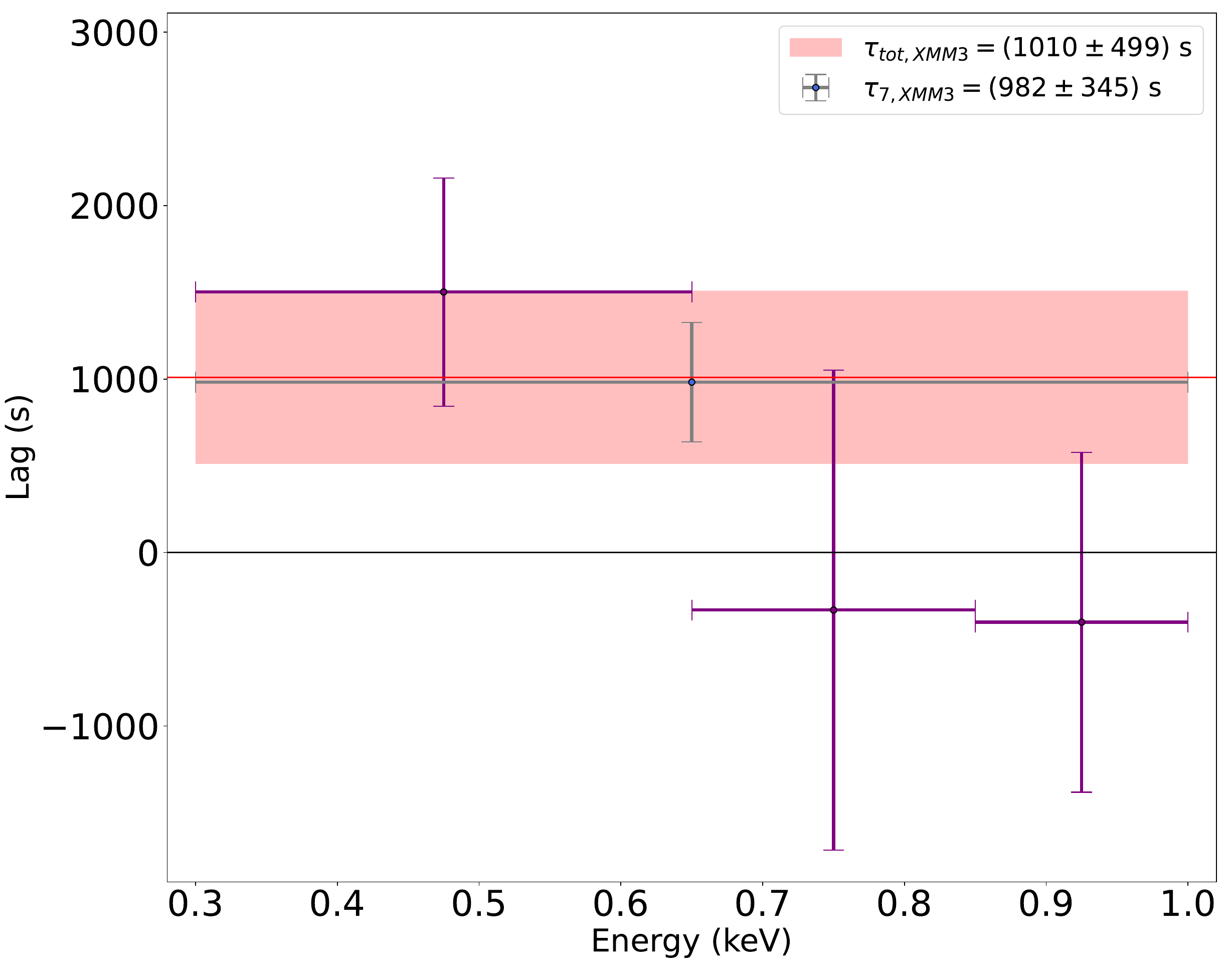}
    \caption{
    Energy-resolved time lags for XMM3-7 (60--70 ks), computed with the same
    setup as in Fig.~\ref{fig:energy_resolved_lag_XMM2}. The gray point
    corresponds to the time window showing the strongest correlation signal
    (see Fig.~\ref{fig:lc_cross_corr_xmm3}).
    }
    \label{fig:energy_resolved_lag_XMM3}
\end{figure}

\label{spec_analysis}

\subsection{Energy Resolved Time lags}
\begin{figure*}[t!]
    \centering
    \begin{minipage}[c]{12cm}
        \centering
        \includegraphics[width=\linewidth]
        {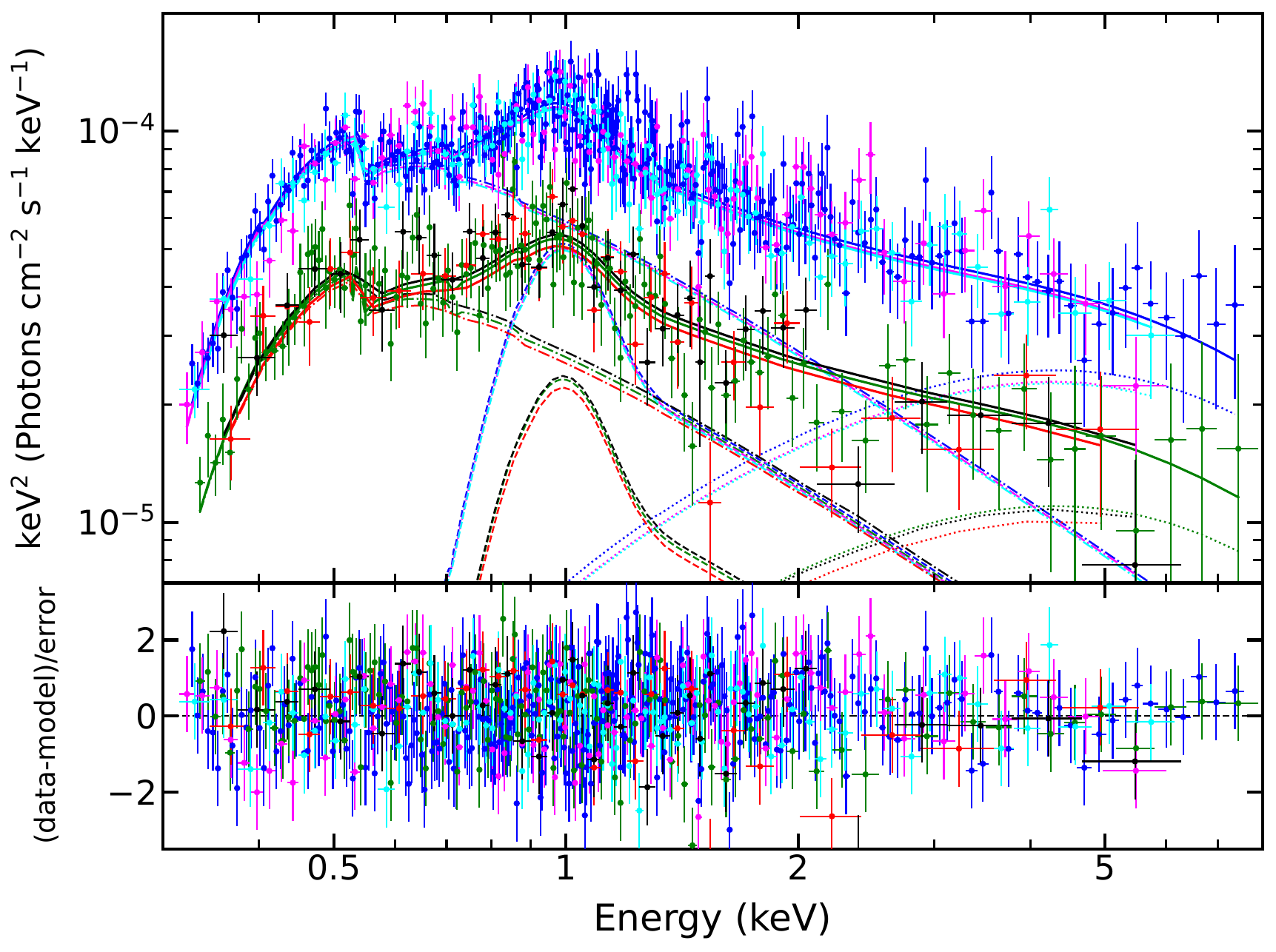}
    \end{minipage}
    \hfill
    \begin{minipage}[c]{\dimexpr\textwidth-12cm-12pt\relax}
        \caption{
             ULX-1 EPIC-pn and MOS (1,2) spectra for XMM2
        (black, red, green) and XMM3 (blue, cyan, purple).
        Dashed lines indicate the best-fit
        {\sc tbabs*thcomp*\\(gauss + bbodyrad + diskbb)}
        model discussed in Section \ref{section:spec_analysis}.
        Best-fit parameters are listed in Table~\ref{tab:thcomp}.
        }
        \label{fig:thcomp_model}
    \end{minipage}
\end{figure*}

\begin{table*}[t!]
\centering
\caption{Best-fit parameters of the Comptonization model discussed in Section \ref{section:spec_analysis}.}

\small
\renewcommand{\arraystretch}{1.1}
\setlength{\tabcolsep}{4pt}

\resizebox{\textwidth}{!}{%
\begin{tabular}{ccccccccccccccc}
\hline
Obs. &
$N_{\rm H}$ &
kT$_{\rm e}$ &
$\Gamma_\tau$ &
kT$_{\rm bbrad}$ &
$N_{\rm bbrad}$ & 
kT$_{\rm diskbb}$ &
$N_{\rm diskbb}$ & 
E$_{\rm line}$ &
$\sigma_{\rm line}$ &
$N_{\rm line}$ &
Flux$^{*}$ & 
$L_X^{**}$ &
$\chi^2_\nu-\mathrm{dof}$ \\
 & [$10^{20}$ cm$^{-2}$] & [keV] & &
 [keV] & & [keV] & [$10^{-4}$] & [keV] & [keV] & [$10^{-5}$] & 
[$10^{-13}$ erg cm$^{-2}$ s$^{-1}$] & [$10^{39}$ erg s$^{-1}$] & \\
\hline

XMM2 &
$9.3^{+4.5}_{-4.6}$ &
\multirow{2}{*}{$\gtrsim 3$} &
\multirow{2}{*}{$3.55^{+0.52}_{-0.74}$} &
$0.087^{+0.033}_{-0.036}$ 
& $297^{+2667}_{-166}$  
& $2.06^{+4.5}_{-0.79}$ 
& $0.95^{+6.7}_{-0.89}$ 
& $0.918^{+0.043}_{-0.052}$
& $0.109^{+0.055}_{-0.040}$
& $1.77^{+0.74}_{-0.30}$ 
& $1.5 \pm 0.03$ 
& $4.3 \pm 0.1$  
& \multirow{2}{*}{0.974-712} \\

XMM3 & 
$9.9^{+3.7}_{-3.5}$ & 
& 
& $0.088^{+0.018}_{-0.018}$ &
$574^{+1123}_{-388}$ &
$1.20^{+0.92}_{-0.18}$ &
$20^{+29}_{-18}$ &
$0.942^{+0.026}_{-0.034}$ &
$0.121^{+0.041}_{-0.046}$ &
$3.41^{+0.83}_{-0.62}$ &
$3.3 \pm 0.03$ & $9.4 \pm 0.1$ &
\\
\hline
\end{tabular}%
}
\label{tab:thcomp}
\tablefoot{
Errors are quoted at the 90\% confidence level.
$^{*}$ EPIC absorbed flux in the 0.3--10~keV energy band.
$^{**}$ EPIC absorbed luminosity in the 0.3--10~keV energy band.
}
\end{table*}
We investigated the contribution of the soft X-ray band to the observed hard lags in the two time windows where the lags are most significant (i.e., 0--10~ks for XMM2 and 60--70~ks for XMM3). We constructed energy-resolved light curves by slicing the soft X-ray data into multiple energy bins and computed the corresponding delays relative to the hard X-ray band.
Both the XMM2 and XMM3 observations indicate that a significant lag is present between the hard X-ray emission and the lowest-energy soft X-ray photons. In particular, Fig. \ref{fig:energy_resolved_lag_XMM2} and Fig. \ref{fig:energy_resolved_lag_XMM3} show that in XMM2 a significant hard lag is detected between the 1.0–10.0 keV band and the 0.3–0.5 keV band, with
\(\tau_{\mathrm{XMM2,1}} \rm [0.3-0.5 \,\,keV] = 2225 \pm 796~\mathrm{s}\).
For XMM3, the hard lag is measured with respect to the 0.3–0.65 keV band, yielding
\(\tau_{\mathrm{XMM3,7}} \rm [0.3-0.65 \,\,keV] = 1502 \pm 658~\mathrm{s}\).

We note that the soft sub-bands used for the energy-resolved lag analysis differ between the two observations. Photon statistics drove this choice: in XMM3, restricting the soft band to 0.3--0.5 keV yielded too few counts to constrain the lag robustly. We therefore slightly extended the lower-energy band to 0.3--0.65 keV, preserving sensitivity to the lowest-energy soft photons while maintaining enough statistics for a meaningful CCF measurement.

\section{Spectral analysis}
\label{section:spec_analysis}
Although ULX-1 is characterized by a remarkably high short-term flux variability \citep[see Section~\ref{short-term_variability} and ][]{pintore20}, the spectral variability is quite suppressed (Fig.~\ref{lcall}, bottom panel) and therefore we extracted the average spectrum from XMM2 and XMM3. 

In the following, we present the results of a physically motivated spectral modeling, in order to explain the energy-resolved time-lag phenomenology shown in Fig. \ref{fig:energy_resolved_lag_XMM2} and Fig. \ref{fig:energy_resolved_lag_XMM3}. The results of alternative spectral modeling performed in this work are reported in the Appendix. We considered a scenario in which the observed emission arises from Comptonization of thermal seed photons in an optically thick medium. To this aim, we adopted the XSPEC model {\sc tbabs*thcomp*(gauss + bbodyrad + diskbb)}, shown in Fig.~\ref{fig:thcomp_model}, and the best-fit parameters are reported in Table~\ref{tab:thcomp}, while a representative physical scenario is depicted in Fig.~\ref{fig:ULX_scenario}. 

In this scenario, the spectrum is interpreted as the superposition of two distinct thermal regions. The outer, cooler region of the accretion flow, likely associated with the outer disk and/or the wind photosphere, is modeled with a {\sc bbodyrad} component. The inner, hotter region, located closer to the compact object, is instead described by a {\sc diskbb} component. Although both thermal components are expected to originate from different regions of the accretion flow and could therefore, in principle, be described by independent {\sc diskbb} models, the limited photon statistics lead to strong degeneracies among the parameters of a two-{\sc diskbb} model, preventing a robust determination of the temperatures and emitting radii of both components.

In both cases, most seed photons from these thermal regions are not observed directly. Instead, they are assumed to undergo Comptonization in a warm, optically thick medium, modeled through the convolution component {\sc thcomp}. This implies that the observed X-ray emission is dominated by reprocessed radiation, rather than by direct thermal emission from the disk. However, we cannot rule out the possibility that a fraction of the seed photons from both thermal components may reach the observer without being fully reprocessed. In particular, unless the system is viewed at extreme inclinations, it is difficult to invoke a geometry in which the Comptonizing medium completely obscures all emitting regions, but rather, a partial leakage of direct thermal photons is expected.
All parameters in Table \ref{tab:thcomp} were left free to vary independently, yielding a very good fit with $\chi^2_\nu = 0.974$ for 712 degrees of freedom.

The results indicate that the soft X-ray emission ($\rm kT_{bbrad}\approx 0.1\,keV$), associated with the {\sc bbodyrad} component (bbrad), originates from a relatively extended region with an internal characteristic radius $R_{\rm bbrad} \approx 4\times10^{4}$~km (inferred from the greatest $N_{bbrad}$ of Table \ref{tab:thcomp}, for an average inclination angle of $60^\degree$). This component is interpreted as arising from the outer, cooler region of the accretion flow (i.e., the blue region of Fig. \ref{fig:ULX_scenario}), possibly linked to the wind photosphere or the outer disk.
The harder component ($\rm kT_{diskbb}\approx 2\,keV$), described by the 
{\sc diskbb} model, is instead associated with a much more compact region, 
with a characteristic inner radius $R_{\rm diskbb} \approx 16$ km, as inferred 
from the lowest $N_{\rm diskbb}$ in Table \ref{tab:thcomp}. 
We stress that this value represents an 
indicative estimate of the characteristic inner radius rather than a precise 
measurement. This value is computed by assuming the lowest {\sc diskbb}  normalization reported in Table~\ref{tab:thcomp} and is subject to substantial uncertainty, owing to both the large statistical uncertainty on the normalization and its 
dependence on the adopted spectral decomposition.
This relatively small characteristic radius is consistent with emission from 
the inner, hotter accretion flow (i.e., the red region of 
Fig. \ref{fig:ULX_scenario}), close to the compact object.
The best-fit parameters, with an electron temperature $kT_e \gtrsim 3$ keV and a photon index $\Gamma_\tau = 3.55^{+0.52}_{-0.74}$, are consistent with a warm, optically thick Comptonizing medium, as expected in the presence of a dense radiatively-driven wind in super-Eddington accretion regimes.
We note that, to test the scenario proposed in Fig. \ref{fig:ULX_scenario}, we attempted to fix the covering fraction of the inner component to unity, assuming it to be almost fully embedded within the wind, while leaving the outer component free to vary, as it is expected to be only partially affected by scattering in the wind photosphere. 
However, due to limited photon statistics, it was not possible to robustly constrain all model parameters. Nonetheless, we find tentative evidence that the lower bound of the covering fraction of the outer component is $\sim 0.33$ (90\% C.L.), partially supporting the proposed scenario.

\begin{figure}[t!]
    \centering
    \includegraphics[width=\linewidth]{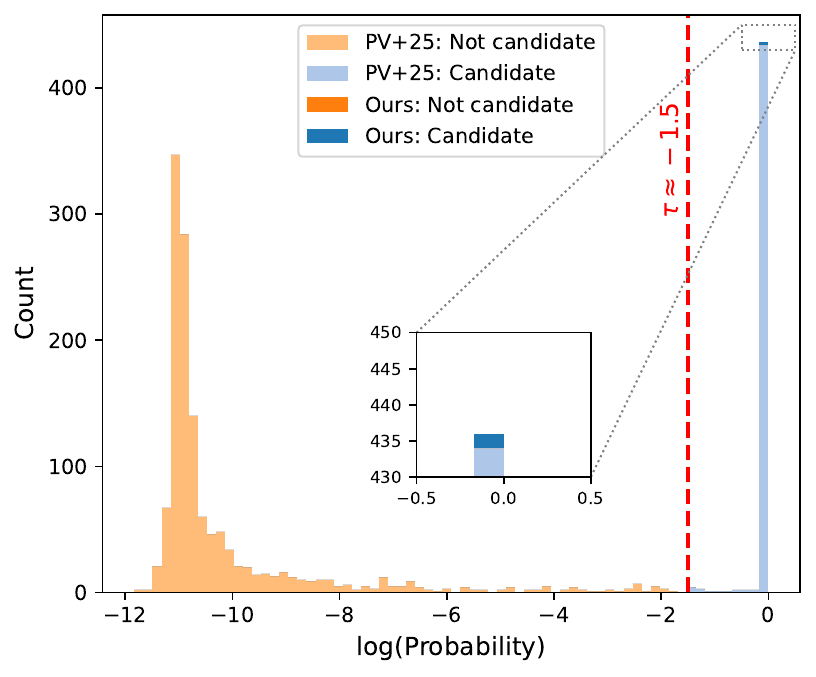}
    \caption{Probability distribution from the Gaussian Mixture Model (GMM) of \citet{pinciroli2025hunt}, applied to the ULX data. The bars are colored according to the classification indicated in \citet{pinciroli2025hunt}: light orange for sources not classified as PULX candidates (``Not candidate''), light blue for PULX candidates (``Candidate''), and dark blue for the two NGC\,7456 ULX-1 observations analyzed here. The latter fall at $\log(\mathrm{Probability}) \approx 0$, placing them within the PULX-candidate cluster.}
    \label{fig:gmm_probability}
\end{figure}

We detect a broad residual structure around $\sim1$ keV, which, in the reference model shown in Fig. \ref{fig:thcomp_model}, is phenomenologically described by a {\sc GAUSS} component convolved with {\sc thcomp}. \footnote{The {\sc thcomp} model describes thermal Comptonization, namely the up-scattering of low-energy seed photons by a population of hot electrons. It computes the resulting Comptonized spectrum self-consistently from the properties of the seed-photon distribution and the scattering medium \citep{Zdziarski_2020}.} Similar features have been reported in several ULXs and are commonly interpreted as blended Fe–L and Ne emission produced in a dense, ionized medium \citep[e.g.,][]{middleton15b,pinto16}. The apparent width of the feature may indicate strong reprocessing, either through repeated scattering or velocity broadening in an optically thick outflow, although the limited EPIC spectral resolution prevents us from robustly constraining its intrinsic width. This is consistent with the presence of powerful winds expected in super-Eddington accretion flows, which can both broaden spectral features and shape the observed X-ray spectrum. By adopting only the continuum model {\sc tbabs*thcomp*(bbodyrad + diskbb)}, the fit yields
$\chi^2/{\rm dof}=850.0/715$. Adding the Gaussian component around
$\sim1$ keV improves the fit to $\chi^2/{\rm dof}=693.5/712$,
corresponding to $\Delta\chi^2=156.5$ for three additional free parameters.
This substantial improvement indicates that the inclusion of an additional
spectral feature around $\sim1$ keV is statistically significant. Indeed, in \citep{pinto23}, the authors show that $\Delta \chi^2$ values of $\sim$40–50 already correspond to a confidence level of approximately 5$\sigma$.
We stress, however, that the phenomenological description of this residual structure is
not unique. Alternative combinations of emission and absorption features can
provide statistically comparable fits (see 
Appendix \ref{appendix:magnetic model} and 
Appendix \ref{appendix:partial covering model}). 
Notably, \citet{pintore20} reported a similar residual structure in 
NGC\,7456 ULX-1, which could alternatively be described as an absorption 
feature at $\sim0.6$--$0.7$ keV or as emission around $\sim1$ keV. Here, we adopt a
single broad emission feature as a physically motivated phenomenological
description, consistent with the presence of reprocessing material also
suggested by the independent timing analysis.

We note significant changes in the source's flux variability between the two observations, as evidenced by variations in the temporal derivative of the count rate, which can be interpreted as fluctuations in the mass accretion rate. These changes are reflected in the different normalizations of the spectral components between the two epochs (see Table \ref{tab:thcomp}), supporting a scenario in which variations in the accretion flow drive the observed spectral changes.

\section{PULX candidate search}
\label{section:pulsation_search}
To assess the nature of the compact object in NGC\,7456~ULX-1 we applied the unsupervised machine-learning pipeline described in  \citet{pinciroli2025hunt} to the available 2018 and 2023 \xmm\ observations of the source. An extensive description of the algorithm is in Appendix \ref{appendix:pipeline}.

Applied to NGC\,7456~ULX-1, the algorithm assigns the source to the PULX-candidate cluster ($C_P$) with a $z$-score $z>1$, robustly above the classification threshold $\tau$ (Fig.~\ref{fig:gmm_probability}). This independently supports the neutron-star accretor hypothesis, placing NGC\,7456~ULX-1 together with the candidate PULXs identified by \citet{pinciroli2025hunt} from the full \xmm\ archive.

Motivated by this classification, we also performed a dedicated search for coherent pulsations in the $(P,\dot{P})$ parameter space with the software Stingray \citep{stingray_2025_17440371}. We identified a candidate signal at a frequency of $\sim0.227$ Hz, corresponding to a spin period of $\sim4.4$ s. However, given the limited statistical significance after accounting for the explored parameter space and the absence of an independent confirmation, we conservatively refrain from claiming a detection and leave it to be confirmed in future observations. 

\section{Discussion}
\label{discussion}

In this paper, we presented a combined spectral and timing analysis of NGC~7456~ULX-1. Using the timing techniques introduced in \citet{Leone2025}, we detect significant hard lags of order $\sim 10^{3}$ s, with the 1--10 keV emission delayed with respect to the 0.3--1 keV band in both the \xmm\ observations considered here. These delays are much longer than the hard lags typically observed in Eddington-limited X-ray binaries, where they are generally of the order of sub-seconds and are commonly interpreted in terms of inward propagation of mass-accretion-rate fluctuations through the accretion flow \citep{nowak99, Kotov_2001}. Conversely, hard lags reaching kilosecond timescales have been observed in active galactic nuclei \citep{kara14}, even if lags of comparable magnitude have been reported in a small number of ULXs \citep[e.g.][]{Zoghbi_2010, hernandez-garcia15}.

By resolving the lag as a function of time and energy, we find that: (i) the hard lag persists across the observations and is particularly significant during phases of strong variability in the count-rate derivative, as shown in Fig. \ref{fig:lc_cross_corr_xmm2} anf Fig. \ref{fig:lc_cross_corr_xmm3}, and (ii) it is primarily driven by the lowest-energy bands, namely 0.3--0.5 keV in the first observation and 0.3--0.65 keV in the second, i.e. the delay is measured between the 1--10 keV emission and these softer bands (see Fig. \ref{fig:energy_resolved_lag_XMM2} and Fig. \ref{fig:energy_resolved_lag_XMM3}). These results suggest a physical connection between the observed delays and the propagation of fluctuations within the accretion flow as proposed in \citet{Zoghbi_2010, hernandez-garcia15}.

\begin{figure}[t!]
    \centering
    \includegraphics[width=0.99\linewidth]{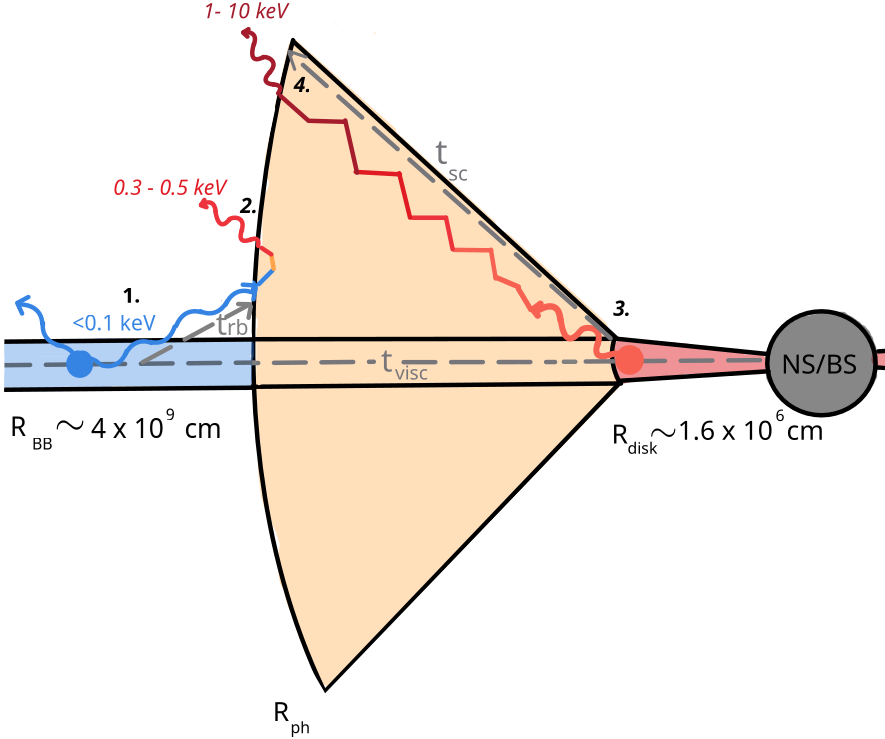}
    \caption{Schematic view of the proposed ULX geometry and lag origin. Wavy colored lines indicate emission at different energies. Gray dashed lines indicate the viscous propagation timescale of the accreted matter and the characteristic scattering timescales associated with the step-like paths, illustrating multiple photon scatterings. (1) Variations in the mass accretion rate ($\dot{m}$) are assumed to originate in the outer disk, enhancing the (\texttt{bbodyrad}) soft emission ($<0.1$ keV). (2) After a reverberation timescale ($t_{\rm rb}$), the {\sc bbodyrad} emission propagates out to the wind photospheric radius ($R_{\rm ph}$), where it undergoes multiple scatterings in the outer optically thick medium and eventually escapes in the $\sim 0.3$--$0.5$ keV energy range. We note that the energy ranges indicated in the figure are representative of the \xmm\ observing band. In particular, the apparent lower boundary at $\sim 0.3$ keV reflects the instrumental sensitivity limit. The underlying emission and reprocessing are expected to extend to lower energies, which are not directly probed in this work. (3) After a viscous timescale $t_{\rm visc}$, the accretion flow propagates inward to the inner disk, where it is expected to enhance the hotter \texttt{diskbb} emission component at $\sim 1.6$ keV; the photons may then undergo further Comptonization and escape after a timescale $t_{\rm sc}$. (4) The resulting hard emission is observed in the 1–10 keV range.}
    \label{fig:ULX_scenario}
\end{figure}

From a spectral perspective, the emission is well described by a Comptonization-dominated model, in which the observed radiation is not the direct emission from thermal seed photons, but rather their reprocessed counterpart. In this framework, the seed photons originate from two distinct thermal regions: (i) a cooler ($kT_{\rm BB} \sim 0.09$ keV, $R_{\rm BB}\sim4\times10^9$ cm), more extended component ({\sc bbodyrad}), associated with the outer disk; (ii) a hotter ($kT_{\rm disk} \sim 2$ keV, $R_{\rm disk}\sim1.6\times10^6$ cm), compact region ({\sc diskbb}) tracing the innermost accretion flow.
According to the model in Fig. \ref{fig:ULX_scenario}, the inner region is almost completely covered by an extended, optically thick medium, so that its emission is entirely reprocessed through multiple Compton scatterings. The outer region likely lies beyond the wind photosphere, and only the fraction of its emission interacting with the optically thick medium is significantly Comptonized and shifted into the \xmm\ energy range \citep{poutanen07,middleton15b}.
As a result, the observed X-ray emission consists of two reprocessed components, both of which emerge at higher energies than the original seed photons.
We note that the accretion flow, and its associated thermal emission, should not be interpreted as being sharply confined in the wind boundaries. Rather, the two seed-photon regions may extend into the wind/photospheric region. In this framework, the inferred scales $R_{\rm BB}$ and $R_{\rm disk}$ should be regarded, respectively, as lower limits to the directly visible emitting regions, while the disk may continue further inward. However, emission from these inner radii would be hidden and reprocessed by the optically thick wind, since $\tau \gg 1$, and therefore would not be directly observable.

The detection of a spectral feature at $\sim 1$ keV, possibly associated with unresolved Fe--L/Ne transitions, further supports the presence of ionized material in the system. However, given the limited EPIC spectral resolution in the soft X-ray band, its apparent width should not be over-interpreted as intrinsic broadening. This is also consistent with the emerging picture from high-resolution X-ray spectroscopy, where accretion-disk winds appear structured and clumpy rather than smooth, as predicted by super-Eddington simulations \citep[][]{Takeuchi_2013,kobayashi18}.

Within the framework shown in Fig. \ref{fig:ULX_scenario}, the observed hard lag can be interpreted as the result of inward propagation of accretion-rate fluctuations combined with radiative diffusion. Fluctuations are expected to originate in the outermost regions of the disk and propagate inward on the viscous timescale \\ \citep{Mahmoud_2018}. As they reach the outer photospheric region, they enhance the {\sc bbodyrad} component, whose seed photons are reprocessed in the wind photosphere and emerge at soft X-ray energies. Consistently, the energy-resolved lag analysis in Fig. \ref{fig:energy_resolved_lag_XMM2} and Fig. \ref{fig:energy_resolved_lag_XMM3} shows that the delays are measured with respect to the 0.3–0.5 keV and 0.3–0.65 keV bands, where the soft emission dominates. We note that the slightly different soft bands adopted for XMM2 and XMM3 reflect the different photon statistics of the two datasets, while both bands probe the lowest-energy emission accessible with \xmm. We stress that this energy range corresponds to the lowest energies accessible with \xmm\, and therefore to the portion of the soft emission where the lag is observationally constrained in this work.
The same perturbations continue to propagate inward, eventually reaching the inner disk and enhancing the {\sc diskbb} component, after a timescale $t_{\rm visc}$. Here, the emission is hotter ($\sim 2$ keV), and photons undergo multiple scatterings within a denser, broader medium, increasing the effective diffusion time ($t_{\rm sc}$) and producing the delayed hard X-ray emission. Therefore the measured lag can be expressed as:
\begin{equation}
\tau \approx t_{\rm visc} + t_{\rm sc}.
\label{eq:tot_delay}
\end{equation}

This would explain both the magnitude of the lag ($\sim 10^{3}$ s) and its stability, as well as the fact that the soft emission precedes the hard one. Moreover, the scenario depicted in Figure~\ref{fig:ULX_scenario} naturally accounts for the typical spectral variability observed in ULXs, with stronger variations in the intermediate energy range ($\sim 1$--$7$ keV) and a tendency to converge at higher energies \citep{pinto23}, as expected from a stratified accretion flow and Comptonizing medium. Furthermore, we note that alternative scenarios based on simple reverberation \citep[see e.g.][]{uttley14} are unlikely to explain the observed delays. In such cases, hard photons would be reprocessed into softer energies, leading to soft lags on light-crossing timescales, therefore, the opposite of our findings.

On the other hand, the relatively low coherence measured between the soft and hard bands further supports this interpretation. Multiple scatterings within the optically thick medium introduce stochastic delays and redistribute photon energies, reducing the phase coherence between different energy bands and effectively damping intrinsic spectral variability through Comptonization.

The characteristic inner radius inferred from the DISKBB normalization is 
$R_{\rm disk} \sim 16$ km. This value corresponds to the best-fit DISKBB 
normalization reported for XMM2 in Table~2 and should be interpreted with caution, 
since the normalization is affected by large statistical uncertainties and 
depends on the adopted spectral decomposition, as also illustrated by the 
alternative models reported in Appendix \ref{appendix:magnetic model} and Appendix \ref{appendix:partial covering model}. 
If we assume that this 
characteristic radius traces the inner accretion disk and that it is larger than 
the innermost stable circular orbit (ISCO), for a non-rotating compact object, 
the compact-object mass satisfies:
\begin{equation}
M \leq \frac{R_{\rm disk}c^2}{6G}.
\end{equation}
Therefore, the relatively small characteristic radius inferred from the spectral 
modeling disfavors an IMBH or even more massive BH accretor, while it does not 
by itself provide strong evidence for a neutron-star accretor.
We stress that this estimate should not be regarded as a robust upper 
limit on the compact-object mass because of the statistical and model-dependent 
uncertainties affecting the DISKBB normalization.

However, the possibility of NGC\,7456 ULX-1 as a neutron-star ULX is supported by the classification of the source as a PULX candidate by the machine-learning framework of \citet{pinciroli2025hunt}, and the presence of a tentative periodicity at $\sim4.4$ s identified in our timing analysis (see Section \ref{section:pulsation_search}). Although the latter does not meet the statistical requirements for a firm detection, it is consistent with the neutron-star scenario and motivates deeper timing observations. A similar result was recently reported by \citet{Yao2026}, where the authors identified a candidate pulsation at $\sim0.22$ Hz ($\sim4.5$ s) in the 2023 XMM-Newton observation, supporting the possible neutron-star nature of NGC\,7456 ULX-1.

Given the tentative nature of the $\sim4.4$ s periodicity, 
a black-hole accretor cannot be excluded. Importantly, the latter 
would not qualitatively alter our interpretation of the hard lag, since hard 
lags can naturally arise in black-hole accretion flows through the inward 
propagation of accretion-rate fluctuations (see, e.g., 
\citealt{Kotov_2001,uttley05,uttley14}). Nevertheless, the $\sim10^3$ s 
lag measured in NGC\,7456 ULX-1 is longer than the hard lags 
typically observed in Galactic stellar-mass black-hole binaries and is more 
comparable to the timescales observed in AGN, where the larger black-hole 
mass naturally shifts the characteristic variability timescales to larger
values. An interpretation based solely on black-hole mass scaling would 
therefore favor a substantially more massive accretor; however, such a 
scenario is disfavored by the small characteristic inner-disk radius inferred 
from our spectral modeling, as discussed above. We therefore cannot exclude 
a stellar-mass black-hole accretor, particularly since the lag timescale also 
depends on the geometry, propagation timescale, and optical depth of the 
super-Eddington flow, but neither does the observed lag provide specific 
evidence in favor of a black hole. Taken together, the timing and spectral 
properties remain suggestive of a neutron-star accretor, although they do not 
provide an unambiguous identification of the compact object.

Assuming that the accretor is a neutron star, and considering the bolometric luminosity formulation in \cite{Mao2026BeamingULX},
we infer a mass accretion rate of
\begin{equation}
\dot{M} \approx 7 \times 10^{-7}\,\mathrm{M_\odot\,yr^{-1}},
\end{equation}
consistent with the minimum values inferred for pulsating ULXs \citep{King_2017}.

Should the dominant contribution to the observed lag arise from photon diffusion, the diffusion timescale can be estimated as \citep{Sunyaev_1980}:
\begin{equation}
t_{\rm diff} \sim \frac{R^{2}}{\lambda c} = \frac{R^{2}\,\rho\,\sigma_{\rm T}}{m_p c},
\end{equation}
where $\lambda = (n_e \sigma_{\rm T})^{-1}$ is the photon mean free path.  This corresponds to an effective density $\rho_{\rm wind}$:
\begin{equation}
\rho_{\rm wind} \sim 7.5\times10^{-5}
\left(\frac{\tau_{\rm obs}}{10^3\,{\rm s}}\right)
\left(\frac{R_{\rm wind}}{10^9\,{\rm cm}}\right)^{-2}
{\rm g\,cm^{-3}},
\end{equation}
where $\tau_{\rm obs}$ is the observed hard lag and $R_{\rm wind}$ represents 
the characteristic radial extent of the wind, from its launching region to 
the outer photospheric radius $R_{\rm ph}$ that must satisfy the condition $R_{ph}<R_{BBrad}$. This scaling provides an effective 
density estimate for a generic compact-object accretor and does not depend 
explicitly on whether the central object is a neutron star or a black hole. 
Such an extended, radiation-driven outflow geometry is broadly consistent 
with super-Eddington accretion simulations (e.g. \citealt{Takeuchi_2013}).
This value should not be interpreted as the average density of a spherical outflow, which would imply unrealistically large mass-loss rates, but rather as an effective local density in the optically thick inner wind or funnel wall where the photons undergo multiple scatterings. In a neutron-star ULX, such dense material could be supplied by a strongly super-Eddington disk, with radiation-pressure-driven outflows launched around the spherization radius and possibly further compressed or structured by the disk geometry, clumpiness, and the interaction with the magnetosphere.

Future simultaneous Optical/UV/X-Ray observations will be crucial for probing the outer, cooler regions of the accretion flow, where seed fluctuations are expected to originate, thereby providing an independent test of this scenario. Indeed, according to the model described in Fig.~\ref{fig:ULX_scenario}, an additional external hard lag is expected. In particular, photons produced in the outermost disk regions at energies $\lesssim0.1$ keV, corresponding to region (1), would precede the soft X-ray photons observed in the 0.3-0.5 keV band, which are produced when the same low-energy photons are scattered and reprocessed at the wind photosphere, corresponding to region (2).
Detecting such a lag would provide a direct probe of the wind photospheric extent, since the delay between the two emission components can be approximated as
\begin{equation}
t_{\rm rb} \approx \frac{R_{\rm BBrad} - R_{\rm ph}}{c},
\end{equation}
where $R_{\rm BBrad}$ is the inner radius of the cooler outer-disk emitting region, and $R_{\rm ph}$ is the characteristic radius of the wind photosphere. Adopting photospheric radii comparable to those found in the super-Eddington simulations of \citet{Takeuchi_2013}, we predict $t_{\rm rb}$ to be of the order of a fraction of a second.

\section{Conclusions}

We have presented a combined spectral and timing analysis of NGC~7456~ULX-1 based on the two deepest available \xmm\ observations, obtained in 2018 and 2023. Using adaptive-binning cross-correlation techniques optimized for the low-count regime, we detected significant hard X-ray lags of order $\sim10^{3}$ s, with the 1--10 keV emission delayed relative to the 0.3--1 keV band.

The average lags measured over the full observations are $\tau_{\rm XMM2}=1230\pm460$ s and $\tau_{\rm XMM3}=1010\pm500$ s. Time-resolved analysis reveals that the most significant episodes occur during intervals with pronounced variations in the time derivative of the count rate, suggesting a connection between the measured delays and rapid changes in the accretion flow. The lag is also energy dependent, being mainly driven by the lowest-energy soft photons: the 1--10 keV emission is delayed with respect to the 0.3--0.5 keV band in XMM2 and to the 0.3--0.65 keV band in XMM3, with no comparable significant delay detected from the higher-energy part of the soft band.

From the spectral point of view, the source is well described by a Comptonization-dominated model in which the observed X-ray emission is largely reprocessed in a warm, optically thick medium. The seed photons are associated with two thermal regions: a cooler, extended component with $kT_{\rm BB}\sim0.1$ keV and $R_{\rm BB}\sim4\times10^{9}$ cm, and a hotter, compact component with $kT_{\rm disk}\sim2$ keV and $R_{\rm disk}\sim1.6\times10^{6}$ cm. These radii should be interpreted as effective scales of the visible or reprocessed emitting regions, since part of the accretion flow may be hidden within the optically thick wind.

The sign, amplitude, temporal behavior, and energy dependence of the measured delays are difficult to reconcile with standard reverberation scenarios. Instead, we interpret the hard lag as the combined effect of the inward propagation of mass-accretion-rate fluctuations and photon diffusion in an optically thick outflow. In this picture, fluctuations first modulate the cooler outer region, whose reprocessed photons emerge at the lowest observable soft X-ray energies, and later reach the hotter inner flow, where photons undergo more extended Compton scattering before escaping in the 1--10 keV band.

By comparing the observed delays with photon-diffusion timescales, we derive the characteristic wind-density scaling as a function of the spatial scale of the wind and the observed hard lag, consistent with the presence of an optically thick outflow expected in super-Eddington accretion. The relatively small characteristic radius inferred from the 
spectral modeling disfavors an IMBH or even more massive BH accretor, although the large statistical and model-dependent uncertainties prevent us from placing a robust constraint on the compact-object mass. Together with the timing properties, these results are suggestive of a neutron-star accretor, although a stellar-mass black hole cannot be excluded.

Overall, our results support a scenario in which the spectral and timing properties of NGC~7456~ULX-1 are shaped by the coupling between propagation-driven variability and radiative transfer in an optically thick wind. Extending this analysis to a larger ULX sample will help establish whether similar hard lags are a common feature of super-Eddington accretion flows.

\begin{acknowledgements} 
This work has been supported by Fondazione di Sardegna through the project “SMART-TIME: AI-Driven Timing and Radio Variability for Pulsars, GRBs \& Sgr A”.
This work has been partially supported by the ASI-INAF program I/004/11/4 and program I/004/11/6. FP acknowledges support from INAF grants OBIWAN (Observing high B-fIeld Whispers from Accreting Neutron stars, 1.05.23.05.12) and BLOSSOM (1.05.23.01.13). FP, CP, GR, AW, RS and CS acknowledge financial support from the Bando Ricerca Fondamentale INAF 2024 Large Grant "Timing the Ultra Luminous x-ray Pulsars (TULiP)". FB acknowledges support from  INAF Large Grant 2023 BLOSSOM F.O. 1.05.23.01.13. S.B acknowledges funding from PRIN MUR 2022 SEAWIND 2022Y2T94C, supported by European Union - Next Generation EU, Mission 4 Component 1 CUP C53D23001330006 and INAF Large Grant 2023 BLOSSOM O.F. 1.05.23.01.13.
\end{acknowledgements}

\section*{Data availability}
All of the data underlying this article are already publicly available from ESA's \xmm\ Science Archive (XSA, \url{https://www.cosmos.esa.int/web/xmm-newton/xsa}), NASA's HEASARC archive (\url{https://heasarc.gsfc.nasa.gov/}).

\addcontentsline{toc}{section}{Bibliography}
\bibliographystyle{aa}
\bibliography{biblio}
\appendix

\label{lastpage}

\section{Adaptive-binning timing technique}
\label{appendix:adaptive_binning}

The time-lag analysis presented in this work relies on the adaptive-binning timing framework introduced by \citet{Leone2025}, specifically designed to estimate delays between light curves in the low-count regime, where standard cross-correlation techniques based on fixed time bins may be strongly affected by Poisson fluctuations.

The method constructs light curves using adaptive binning with a fixed number of counts per bin. In practice, the temporal bin edges are chosen so that each bin contains the same number of photons (n), rather than having a fixed duration. As a result, the time resolution automatically adapts to the instantaneous source brightness, providing finer sampling during high-flux intervals and larger bins when the count rate decreases.

For each energy band, the arrival times of the events are grouped into adaptive bins independently, and the count rate associated with each bin is computed as:
\begin{equation}
    R_i = \frac{n}{\Delta t_i},
\end{equation}
where ($n$) is the fixed number of photons within each interval, and $\Delta t_i$ is the time interval containing $n$ ToAs.

The resulting light curves are then interpolated onto a common temporal grid and cross-correlated using the discrete cross-correlation function (CCF). Following \citet{Leone2025}, the time delay is estimated from the position of the CCF peak. When a single statistically significant peak is identified, the peak region is fitted with a Gaussian profile. The fitting interval is defined once from the CCF of the original data and kept fixed throughout the subsequent Monte Carlo analysis to avoid introducing additional bias from fluctuations in the CCF shape. If no unique significant peak is present, a Gaussian function is instead fitted over the full CCF profile, thereby avoiding subjective choices in the lag determination.

The statistical uncertainty on the lag is evaluated using the Modified Double Pool (MDP) method introduced by \citet{Leone2025}. This approach combines adaptive binning with a bootstrap resampling scheme based on random halving of the photon lists. For each realization, the events in each energy band are randomly divided into two independent subsets, from which synthetic light curves are generated and cross-correlated. The lag measurement is repeated across a large number of realizations while preserving the same fitting interval used for the original CCF. The final delay and its uncertainty are derived from the distribution of the recovered lags. Extensive simulations presented in \citet{Leone2025} show that the MDP method provides unbiased lag estimates and robust confidence intervals in the low-count regime, significantly outperforming standard fixed-binning techniques.

\section{Magnetic model}
\label{appendix:magnetic model}
\begin{figure*}[t!]
    \centering
    \begin{minipage}[c]{12cm}
        \centering
        \includegraphics[width=\linewidth]
        {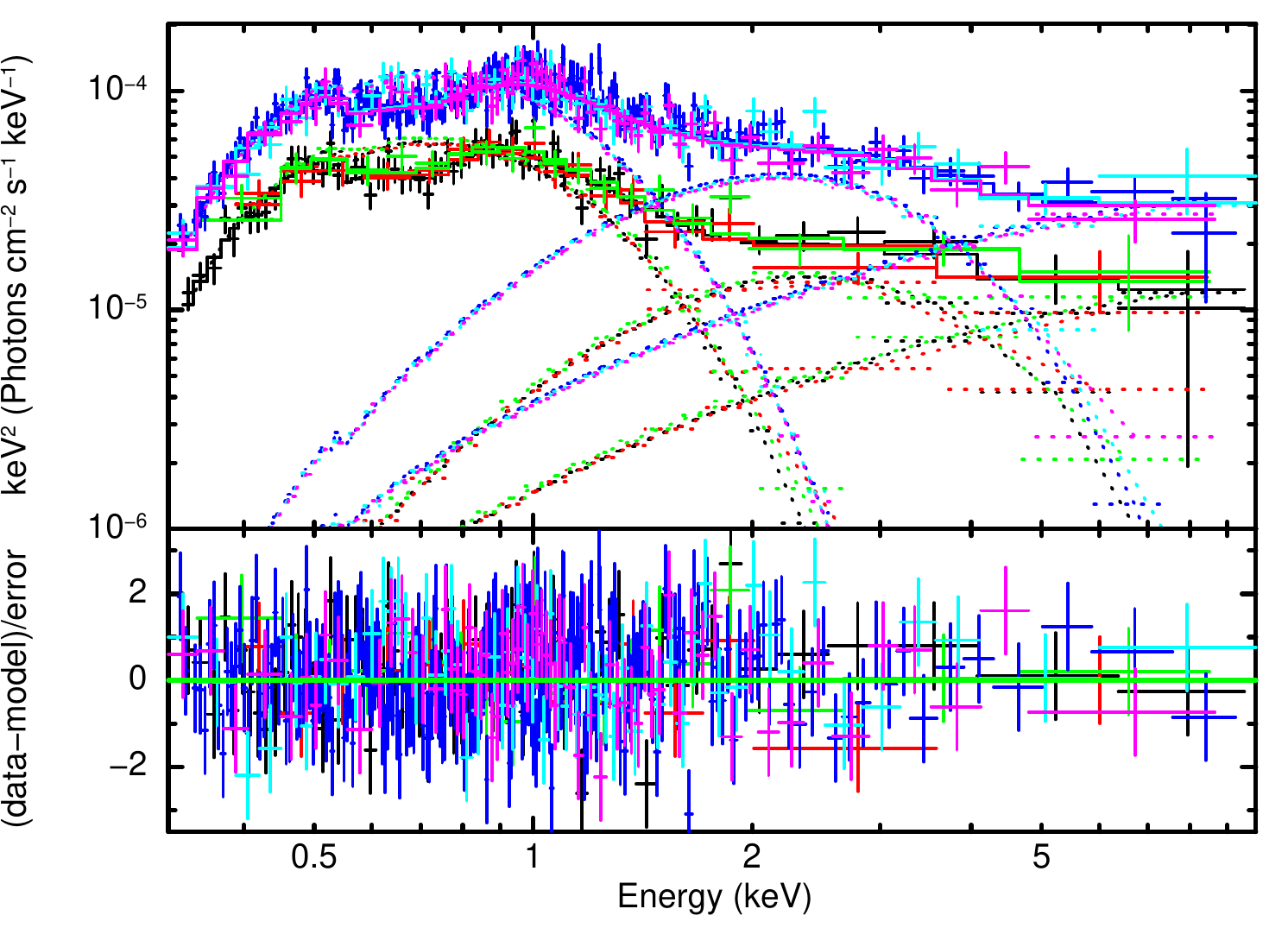}
    \end{minipage}
    \hfill
    \begin{minipage}[c]{\dimexpr\textwidth-12cm-12pt\relax}
        \caption{
            ULX-1 EPIC-pn and MOS (1,2) spectra for XMM2
            (black, red, and green for EPIC-pn, MOS1, and MOS2, respectively)
            and XMM3
            (blue, cyan, and purple for EPIC-pn, MOS1, and MOS2, respectively).
            Dashed lines indicate the best-fit
            \textsc{tbabs (gaussian + diskbb + bbodyrad + cutoffpl)} model.
        }
        \label{spectra_best-fit}
    \end{minipage}
\end{figure*}
\begin{table*}[t!]
\center
\caption{Best-fit spectral parameters for the alternative models discussed in Appendix \ref{appendix:magnetic model} and \ref{appendix:partial covering model}.}
\scalebox{0.6}{\begin{minipage}{31cm}
      \label{spec_par}
\begin{tabular}{ccccccccccccccccc c c}
\hline
Obs. &  nH  & $kT_{\mathrm{dbb}}$ & N$_{\text{dbb}}$ & $kT_{\mathrm{bb}}$ & N$_{\text{bb}}$ & N$_{\text{cutoffpl}}$$^{*}$ & E$_{line}$ & $\sigma_{line}$ & N$_{line}$ & E$_{line}$ & $\sigma_{line}$ & N$_{line}$ & $nH_{pcfabs}$ & $f$ & Flux$^{**}$ & L$_X$$^{***}$ & $\chi^2_{\nu}-\text{dof}$ \\
 &  [$10^{20}$ cm$^{-2}$] & [keV] & & [keV] & [$10^{-3}$] & [$10^{-6}$] & [keV] & [keV] & [$10^{-6}$] & [keV] & [keV] & [$10^{-6}$] & [$10^{22}$ cm$^{-2}$] & & [$10^{-13}$ erg cm$^{-2}$ s$^{-1}$] & [$10^{39}$ erg  s$^{-1}$]  & \\
\hline
XMM2 & \multirow{2}{*}{$6.9_{-1.3}^{+1.5}$} & \multirow{2}{*}{$0.26_{-0.02}^{+0.02}$} & $2.4_{-0.6}^{+1.0}$ & $0.88_{-0.09}^{+0.10}$ & $7_{-2}^{+4}$ & - & $0.68_{-0.04}^{+0.03}$ & $0.07_{-0.03}^{+0.04}$ & $6.5_{-3}^{+5}$ & - & - & - & - & - & $1.40\pm 0.05$ & $4.1\pm0.1$ & \multirow{2}{*}{$1.06-712$} \\

 XMM3 & & & $5.1_{-1.4}^{+2.2}$ & $0.77_{-0.04}^{+0.04}$ & $30_{-6}^{+8}$ & - & $0.66_{-0.04}^{+0.03}$ & $0.10_{-0.03}^{+0.03}$ & $22_{-8}^{+15}$ & - & - & - & - & - & $3.2\pm0.06$ & $9.4\pm0.2$ &  \\

\hline
\hline
XMM2 & \multirow{2}{*}{$11_{-2}^{+3}$} & \multirow{2}{*}{$0.21_{-0.02}^{+0.02}$} & $7.2_{-3}^{+8}$ & $0.62_{-0.13}^{+0.15}$ & $20_{-10}^{+29}$ & $2.0_{-1.5}^{+1.2}$ & $0.63_{-0.08}^{+0.05}$ & $0.11_{-0.05}^{+0.05}$ & $20_{-11}^{+34}$ & - & - & - & - & - & $1.45\pm 0.05$ & $4.3\pm0.1$ & \multirow{2}{*}{$1.00-710$} \\

 XMM3 & & & $15.2_{-7}^{+18}$ & $0.55_{-0.05}^{+0.06}$ & $94_{-32}^{+49}$ & $5.3_{-1.1}^{+1.0}$ & $0.61_{-0.10}^{+0.05}$ & $0.14_{-0.03}^{+0.05}$ & $54_{-27}^{+94}$ & - & - & - & - & - & $3.37\pm0.07$ & $9.9\pm0.2$ &  \\
\hline
%\smallskip
\hline
XMM2 & \multirow{2}{*}{$11_{-2}^{+3}$} & \multirow{2}{*}{$0.20_{-0.02}^{+0.02}$} & \multirow{2}{*}{$24_{-11}^{+28}$} & \multirow{2}{*}{$0.55_{-0.05}^{+0.06}$} & \multirow{2}{*}{$99_{-35}^{+56}$} & \multirow{2}{*}{$5.4_{-1.1}^{+1.0}$} & \multirow{2}{*}{$0.62_{-0.09}^{+0.05}$} & \multirow{2}{*}{$0.13_{-0.03}^{+0.04}$} & \multirow{2}{*}{$67_{-35}^{+116}$} & - & - & - & $>54$ & $0.64_{-0.04}^{+0.04}$ & \multirow{2}{*}{} & \multirow{2}{*}{} & \multirow{2}{*}{$1.01-714$} \\

 XMM3 & & & & & & & & & & - & - & - & $0.71_{-0.44}^{+0.76}$ & $0.27_{-0.07}^{+0.07}$ & & & \\
\hline
\hline
XMM2 & \multirow{2}{*}{$6.2_{-1.4}^{+1.5}$} & \multirow{2}{*}{$0.239_{-0.015}^{+0.016}$} & \multirow{2}{*}{$8_{-2.5}^{+3.6}$} & \multirow{2}{*}{$0.59_{-0.06}^{+0.06}$} & \multirow{2}{*}{$79_{-24}^{+39}$} & \multirow{2}{*}{$5.1_{-1.3}^{+1.1}$} & \multirow{2}{*}{$0.68_{-0.02}^{+0.02}$} & \multirow{2}{*}{0.05} & \multirow{2}{*}{$9.8_{-3.6}^{+4.2}$} & \multirow{2}{*}{$0.99_{-0.02}^{+0.02}$} & \multirow{2}{*}{0.05} & \multirow{2}{*}{$4.7_{-1.5}^{+1.6}$} & $>45$ & $0.65_{-0.04}^{+0.04}$ & \multirow{2}{*}{} & \multirow{2}{*}{} & \multirow{2}{*}{$0.995-712$} \\

 XMM3 & & & & & & & & & & & & & $0.88_{-0.4}^{+0.6}$ & $0.29_{-0.07}^{+0.07}$ & & & \\
 \hline
\end{tabular}
\end{minipage}}
\tablefoot{
The model and spectral analysis are discussed in
Appendix \ref{appendix:magnetic model} and \ref{appendix:partial covering model}.
Errors are quoted at the 90\% confidence level.
$^{**}$ EPIC absorbed flux in the 0.3--10~keV energy band.
$^{***}$ EPIC absorbed luminosity in the 0.3--10~keV energy band.
}
\flushleft $^*$ The parameters of the {\sc cutoffpl} were fixed at $\Gamma=0.59$ and kT$_{e}=7$ keV; \\ $^{**}$ EPIC absorbed flux in the 0.3--10 keV energy band; \\ $^{***}$ 0.3--10 keV EPIC absorbed luminosity;

\end{table*}

\citet{pintore20} showed that the combination of two absorbed ({\sc tbabs} in {\sc Xspec}) thermal components ({\sc diskbb} and {\sc bbodyrad} in {\sc Xspec}) gives a good modeling of the spectral properties of ULX-1, and we initially adopted the same model here. In addition, a {\sc gaussian} absorption line at $\sim0.65$ keV was also included.

We fitted the EPIC-pn and MOS (1 and 2) spectra simultaneously for both observations, allowing all spectral parameters in Table \ref{spec_par} to vary freely. We also included a multiplicative constant to account for calibration differences between instruments. We found that these are compatible within 10$\%$ as expected \citep{madsen15}.
The best-fit models clearly indicate the absence of marked spectral variability, but do show changes in the flux. We found that the absorption column density values are consistent within the uncertainties of the two observations; therefore, we kept them linked. The same applies to the temperature of the {\sc diskbb} component. The final best-fit gives a very good description of the spectra ($\chi^2/dof = 758.11/712 = 1.06$) with temperatures of $\sim0.25$ keV and $\sim0.8$ keV for the {\sc diskbb} and {\sc bbodyrad} components, respectively. The presence of the absorption line is also confirmed, with a central energy of $\sim0.67$ keV. A slight variability in the line spectral normalization is also observed, although it is generally compatible within uncertainties in energy and broadening. In particular, the normalization is stronger during the highest-flux observation (XMM3). We also note a slight variation in temperature for the {\sc bbodyrad} model (0.88 keV vs 0.77 keV for XMM2 and XMM3, respectively). However, the model leaves residuals at energies above 5 keV, suggesting either that the {\sc bbodyrad} component is not broad enough to model these energies or that a third component is needed. We explored the first possibility by substituting the {\sc bbodyrad} component with a {\sc diskbb}. It yields a statistically compatible fit with the previous one ($\chi^2/dof = 730.51/712 = 1.03$), with the hot {\sc diskbb} temperatures converging at $\sim1.3$ keV and $\sim1.1$ keV, although within uncertainties. However, some residuals remain at high energies, indicating the need for an additional spectral component. We therefore add a third non-thermal component ({\sc cutoffpl} in {\sc Xspec}), as is usually done for PULX spectra. We fixed the photon index and the cut-off energy to the average values observed for the PULXs ($\Gamma=0.59$ and kT$_{e}=7$ keV; \citealt{walton18})
and left the normalization free to vary for both observations. The best-fit statistics ($\chi^2/dof = 714.45/710 = 1.01$) indicate that the third component was statistically required ($\Delta\chi^2\sim44$ for 2 additional degrees of freedom). The final best-fit model is reported in Table~\ref{spec_par} and shown in Fig.~\ref{spectra_best-fit}.
The average observed luminosity of the source is $\sim4\times10^{39}$ \lum\ and $\sim9\times10^{39}$ \lum\ in the 0.3--10 keV energy band for XMM2 and XMM3, respectively. Due to the high short-term variability, this implies that the maximum luminosity of ULX-1 is most likely higher than $10^{40}$ \lum\ (assuming a distance of 15.7 Mpc).

\section{Partial covering model}
\label{appendix:partial covering model}
We also adopted an alternative model to describe the average spectra of XMM2 and XMM3. Since no marked spectral variability is observed between the two observations, despite significant flux variability, we investigate the possibility that this behavior is due to partial covering of the emission from an optically thick medium. We adopted the continuum model given by the {\sc (diskbb + bbodyrad + cutoffpl)}, absorbed by {\sc tbabs} and {\sc pcfabs} models, including also the previously reported gaussian absorption line at $\sim0.65$ keV. The final model is thus {\sc tBabs*pcfabs(diskbb + bbodyrad + cutoffpl + gaussian)}. We linked the continuum model, the Gaussian line, and the interstellar medium column density across the spectra of XMM2 and XMM3, and left the column density and the partial-covering factor of {\sc pcfabs} free to vary independently. We obtain a good fit ($\chi^2_\nu = 1.01$ for 713 dof); best-fit parameters are reported in Table~\ref{spec_par}. Continuum and line parameters are all compatible, within uncertainties, with those obtained from the previous analysis. The column density and the covering fraction for XMM2 and XMM3 converged toward $nH_{{\text{{\sc pcfabs}}}} > 54\times10^{22}$ cm$^{-2}$ and $nH_{{\text{{\sc pcfabs}}}} = 0.71_{-0.44}^{+0.76}\times10^{22}$ cm$^{-2}$, and $f = 0.64_{-0.04}^{+0.04}$ and $f = 0.27_{-0.07}^{+0.07}$. The results indicate that an optically thick medium that screens the source's entire emission can be a viable explanation for the absence of spectral variability. We estimated that the unabsorbed 0.3--10 keV flux during XMM2 and XMM3 is $6.5\times10^{-13}$ \flux, which corresponds to a ULX-1 luminosity of $\sim1.9\times10^{40}$ \lum.\\
Finally, we also considered a different modeling of the spectral features by including one absorption and one emission line, replacing the single absorption line at 0.65 keV. The two lines were set at $\sim0.7$ and $\sim1$ keV, with a fixed width of 0.05 keV. The best-fit statistic is compatible with the previous ones. Results are reported in Table~\ref{spec_par}.

\section{Machine-learning pipeline for PULX classification}
\label{appendix:pipeline}
The algorithm is based on a Gaussian Mixture Model (GMM), applied to a database of 640 ULXs comprising 1769 individual \xmm\ observations.  Of these,
95 observations (from 6 confirmed sources) are labeled as known PULXs and are used only to set the decision threshold, while the GMM is fitted exclusively to observations of sources with unknown compact-object nature.  

The GMM represents the full ULX population as the weighted sum of two Gaussian components in the multi-dimensional feature space described in \citet{pinciroli2025hunt}. A positive, large $z$-score corresponds to a high probability of belonging to the PULX-like cluster. The two clusters identified by the GMM are denoted $C_P$ (the cluster containing the majority of known PULXs) and $C_U$ (the remaining sources).  The separation threshold is chosen to maximize the \emph{uncertain ratio} (UR, the fraction of unknown-nature ULXs assigned to $C_U$) subject to the constraint that the \emph{PULX ratio} (PR, the fraction of known-PULX observations assigned to $C_P$) is at least 99\%.  Under this criterion the pipeline correctly recovers $\gtrsim99\%$ of all known PULX observations while isolating 85 new candidate PULXs (355 observations) in $C_P$ \citep{pinciroli2025hunt}.

\end{document}